\documentclass[aps,prb,twocolumn,groupedaddress,showpacs]{revtex4}

\usepackage[bookmarks]{hyperref}
\usepackage[dvips]{graphicx}
\usepackage{amsmath}
\usepackage{amstext}
\usepackage{graphics}
\usepackage{exscale}
\usepackage{epsfig}
\usepackage[T1]{fontenc}
\usepackage{bm}
\usepackage{bbm}

\usepackage{version}

\usepackage{latexsym}

\renewcommand{\epsilon}{\varepsilon}

\newcommand{\ket}[1]{\! | #1\ \!\! \rangle}

\newcommand{\gfbraket}[1]{\langle\!\langle #1  \rangle\!\rangle}

\newcommand{\integral}[3]{\!\int\limits_{#2}^{#3}\!\!{\rm d}#1\;}

\newcommand{\expval}[2]{ \langle  #1 #2\ \!\! \rangle}

\newcommand{\elcre}[2]{ c^{\dagger}_{#1,#2}}
\newcommand{\elann}[2]{ c_{#1,#2}}

\newcommand{\e}{\mathrm e}

\newcommand{\vk}{{\bm k}}
\newcommand{\vq}{{\bm q}}

\newcommand{\thGf}{{\cal G}}

\newcommand{\Imag}{\mathrm{Im}}

\newcommand{\hc}{\mathrm{h.c.}}

\includeversion{commentst1} 

\begin{document}

\title{Competition between antiferromagnetic and charge order in the Hubbard-Holstein model}   
\author{Johannes Bauer${}^1$ and Alex C. Hewson${}^2$}
\affiliation{${}^1$Max-Planck Institute for Solid State Research, Heisenbergstr.1,
  70569 Stuttgart, Germany} 
\affiliation{${}^2$Department of Mathematics, Imperial College, London SW7 2AZ,
  United Kingdom}
\date{\today} 
\begin{abstract}
We study the competition between an instantaneous local Coulomb repulsion and a boson
mediated retarded attraction, as described by the Hubbard-Holstein
model. Restricting to the case of half filling, the ground-state phase diagram and the
transitions from antiferromagnetically ordered states to 
charge ordered states are analyzed. The calculations are based on the 
model in large dimensions, so that dynamical mean field 
theory can be applied, and the associated impurity problem is solved using
the numerical renormalization group method. The transition is found to occur
when  electron-electron coupling strength $U$ and the induced interaction
$\lambda$  due  to electron-phonon coupling  approximately coincide, $U\simeq
\lambda$. We find a continuous transition for small coupling and large
$\omega_0$, and a discontinuous one for large coupling and/or small $\omega_0$. We present results
for the order parameters, the static expectation values for the electrons and
phonons, and the corresponding spectral functions. They illustrate the
different types of  behavior to be seen near the transitions. Additionally,
the quasiparticle properties are calculated in the normal state, which leads
to a consistent interpretation of the low energy excitations.  
\end{abstract}

\maketitle

\section{Introduction}

A feature of  strongly correlated systems is the existence of
competing interactions on low energy scales which can lead to different types
of symmetry breaking and different ground states. 
There can be various forms of magnetic  order,
superconducting or charge ordered states; there may also be
transitions between these states, and in some cases they even coexist.  
For instance, compounds such as the vanadites \cite{IFT98}, high $T_c$
cuprates \cite{DSH03}, fullerides \cite{Gun97}, manganites \cite{MLS95,Mil98} and organic salts
\cite{PM06} possess rather involved phase diagrams and, to understand them, an analysis of
the competition between the  different interactions will be important.

Here we study the competing effects between an instantaneous local Coulomb
repulsion and the retarded interaction induced by a coupling to an
optical phonon mode using the Hubbard-Holstein (HH) model. We consider
 the competition between two types of order,
antiferromagnetic (AFM) and charge order (CO), which can occur in the model at half filling.  The emphasis
will be on treating the phonons fully quantum mechanically and 
  in allowing for arbitrary coupling strengths, so 
that  the full interaction parameter regime can be investigated. 
This is possible if we use the infinite dimensional version of  Hubbard-Holstein 
model so that we can apply the dynamical mean field theory (DMFT), which 
 becomes exact in this limit. The numerical renormalization group (NRG) method is
 then used to solve the associated effective impurity problem. This permits one to handle both strong
electron-electron and strong electron-phonon interactions as well as a wide
range of phonon frequencies. We focus on  the ground
state and spectral properties of electrons and phonons at zero temperature. 

The infinite dimensional Hubbard, Holstein and combined HH models have received
considerable attention in the past
\cite{FJS93,FJ94,FJ95,LBV95,Tak96,BZ98,CP99,HG00,MHB02,DM02a,DM02b,CC03,KMOH04,KMH04,KHE05,SCCG05}. For
the pure Holstein case, Freericks et al. \cite{FJS93,FJ94} found instabilities
to charge order and superconductivity by quantum Monte Carlo (QMC) and
iterated perturbation theory for different filling factors. At half filling,
Benedetti and Zeyher \cite{BZ98}, and Hague and D'Ambrumenil \cite{HD08},
investigated the normal state and found a breakdown of Migdal-Eliashberg
theory when a lattice instability develops for 
stronger electron-phonon coupling. The charge ordered ground state and phase
diagram in the adiabatic limit has been analyzed by Ciuchi et al. \cite{CP99}. It
was shown there that the weak and strong coupling CO states are smoothly connected.

For the Hubbard-Holstein model in the absence of long range order,
the phase diagram of the  paramagnetic (PM), bipolaronic (BP) phases
and the metal-insulator (MI) transition has been  established \cite{JPHLC04,KMOH04}.
Another recent study of the model without long range order
deals with the topic of polaron formation \cite{KHE05,SCCG05}
with finite electron density, extending the original work of Holstein
\cite{holsteinmodel} who considered the single electron case only.
The occurrence of superconductivity was studied in
Ref. \onlinecite{FJ95,LBV95,Tak96}. In a two site calculation Takada found
superconductivity with off-site pairing at half filling in a very small
parameter regime in the antiadiabatic region \cite{Tak96}.
The effect of phonons on the quasiparticle
excitations in the presence of  AFM has also been investigated \cite{SGKCC06}.
There have also been extensive calculations for the one dimensional 
version of model \cite{CH05,FJ06,TTCC07,TAA07,HC07,FHJ08}, which we shall comment on briefly later.

Our analysis here of the HH model will extend the earlier work by allowing for
AFM  and CO states, which are the dominant instabilities at half filling. We 
study the transitions between these states. This will give a more complete
picture of the phase diagram and the properties of the model in the ordered
phases. In the regions of the phase diagram with CO, we 
also obtained superconducting solutions, but the CO states were found to have
lower energy. We calculate the static and dynamic properties in both
these types of  broken symmetry phases.  The paper is structured as
follows. In Sec. II we 
specify the formal setup of the HH  model and the DMFT-NRG method. We also
give explicit expressions for the different contributions to the total
energy. The dependence of these on the interactions is discussed in detail in
Sec. IV. Before that in Sec. III, we discuss the global phase diagram, the
order parameters and static quantities and their dependence on $U$, $\lambda$
and $\omega_0$. Sec. V explores  the normal state properties of the HH model
which helps to understand the ground state phase diagram and transition. In
Sec. VI we discuss how the bosonic properties are modified by the
coupling to the electronic system. In Sec. VII we present results for the
electronic and bosonic spectral functions, before concluding in Sec. VIII.

\section{Model and DMFT-NRG setup}

The Hamiltonian for the HH model is given by
\begin{eqnarray}
  \label{hubholham}
  H&=&-\sum_{i,j,{\sigma}}(t_{ij}\elcre i{\sigma}\elann
j{\sigma}+\hc)+U\sum_i\hat n_{i,\uparrow}\hat n_{i,\downarrow} \\
&&+\omega_0\sum_ib_i^{\dagger}b_i+g\sum_i(b_i+b_i^{\dagger})\Big(\sum_{\sigma}\hat
n_{i,\sigma}-1\Big).
\nonumber
\end{eqnarray}
$\elcre i{\sigma}$ creates an electron at lattice site $i$ with spin $\sigma$,
and $b_i^{\dagger}$ a phonon with oscillator frequency $\omega_0$,
$\hat n_{i,\sigma}=\elcre i{\sigma}\elann 
i{\sigma}$. The electrons interact locally with strength $U$, and
their density is coupled to an optical phonon mode with coupling constant
$g$. We have set the ionic mass to $M=1$ in (\ref{hubholham}). The local
oscillator displacement is related to the bosonic operators by  $\hat
x_i=(b_i+b_i^{\dagger})/\sqrt{2\omega_0}$, where $\hbar=1$, and one can
define a characteristic length $x_0=1/\sqrt{\omega_0}$ for the oscillator. 
In Appendix \ref{mfapp} we give the details for a mean field calculation in
the adiabatic limit for this model.

For our calculations we assume a bipartite lattice with $A$ and $B$ sublattice, where
the matrix Green's function can be written in the form    
\begin{equation}
\underline{G}_{\vk,\sigma}(\omega) \!=\!
\frac1{\zeta_{A,\sigma}(\omega)\zeta_{B,\sigma}(\omega) -\epsilon_{\vk}^2}
\! \left(\!\!\!
\begin{array} {cc}
 \zeta_{B,\sigma}(\omega) & \epsilon_{\vk} \\
\epsilon_{\vk} & \zeta_{A,\sigma}(\omega)
\end{array}
\!\!\!\right),
\label{kgf}
\end{equation}
with $\zeta_{\alpha,\sigma}(\omega)=\omega+\mu_{\alpha,\sigma}-\Sigma_{\alpha,\sigma}(\omega)$,
$\alpha=A,B$, and $\vk$-independent self-energy \cite{MV89}.
For commensurate charge order we have $\mu_{A,\sigma}=\mu-h_c$, $\mu_{B,\sigma}=\mu+h_c$ and
$\Sigma_{B,\sigma}(\omega)=Un-\Sigma_{A,\sigma}(-\omega)^*$, with
$n=(n_A+n_B)/2$, $n_{\alpha}=\sum_{\sigma}n_{\alpha,\sigma}$,
$n_{\alpha,\sigma}=\expval{\hat n_{\alpha,\sigma}}{}$. For the
AFM order one has $\mu_{A,\sigma}=\mu-\sigma h_s$, $\mu_{B,\sigma}=\mu+\sigma
h_s$, and the condition
$\Sigma_{B,\sigma}(\omega)=\Sigma_{A,-\sigma}(\omega)$.  We consider solutions
of exclusive AFM or CO, where the symmetry breaking fields vanish, $h_c,h_s\to
0$. 

In the case with symmetry breaking, the effective Weiss field
is a  $2\times2$ matrix $\underline\thGf_{0}^{-1}(t)$.
The DMFT self-consistency equation in this case reads\cite{GKKR96}  
\begin{equation}
  \underline\thGf_{0,\sigma}^{-1}(\omega)=\underline G_{\sigma}(\omega)^{-1}
  +\underline\Sigma_{\sigma}(\omega).
\label{scselfcon}
\end{equation}
The matrix of local lattice Green's functions $\underline
G_{\sigma}(\omega)=1/N\sum_{\vk}\underline{G}_{\vk,\sigma}(\omega)$ is
obtained by integrating over 
the density of states,
$1/N\sum_{\vk}f(\epsilon_{\vk})=\integral{\epsilon}{}{}\rho_0(\epsilon)
f(\epsilon)$. We assume a semi-elliptic DOS,
$\rho_0(\epsilon)=2\sqrt{D^2-\epsilon^2}/\pi  
D^2$ corresponding to a Bethe lattice in all the following calculations.
In the DMFT  this local  Green's function, and the self-energy
are identified with the corresponding quantities for 
an effective impurity model \cite{GKKR96}. One focuses for the calculations
on the properties of the $A$-sublattice. We can take the form of this
impurity model to correspond to an  Anderson-Holstein impurity model \cite{HM02} and
calculations are carried out as detailed, for instance in Ref. \onlinecite{ZPB02,BH07c}.
We solve the effective impurity problem with the numerical renormalization
group \cite{Wil75,BCP08} (NRG) adapted to these cases with symmetry
breaking. The NRG has been shown to be very successful for calculating the local
dynamic response functions, and we use the recent approach\cite{PPA06,WD07}
based on the complete basis set proposed by Anders and Schiller \cite{AS05}. For
the logarithmic discretization parameter we take the value  $\Lambda=1.8$ and
keep about 1000 states at each iteration. The bosonic Hilbert space is
restricted to a maximum of 50 states.  

In the AFM case the $A$-sublattice magnetization,
$\Phi_{\rm afm}=m_A=(n_{A,\uparrow}-n_{A,\downarrow})/2$ serves as an order parameter.
For CO we define  
$\Phi_{\rm co}=(n_A-1)/2$.

To find the ground state of the system we calculate the ground state energy $E_{\rm
  tot}=\expval{H}{}/N$ of the HH Hamiltonian (\ref{hubholham})
in the different phases. This gives generally,
\begin{equation}
  E_{\rm tot}=E_{\rm kin}+E_U+E_g+E_{\rm ph}. 
\label{Etot}
\end{equation}

The first term is the kinetic energy, which reads
\begin{equation}
  E_{\rm kin}=
\sum_{\sigma}\integral{\epsilon_{\vk}}{}{}\rho_0(\epsilon_{\vk})\epsilon_{\vk}\integral{\omega}{}{}f(\omega)\rho_{AB,\vk,\sigma}(\omega),    
\end{equation}
where $\rho_{AB,\vk,\sigma}(\omega)=-\Imag G_{AB,\vk,\sigma}(\omega)/\pi$ for
the off-diagonal Green's function in (\ref{kgf}) and
$f(\omega)$ is the Fermi function, where $f(\omega)=\theta(-\omega)$ at zero
temperature. In the non-interacting case this expression can be evaluated analytically 
and we find for half filling, $\mu=0$, $E_{\rm kin}^0=-4D/3\pi$, which for
$D=2$ is $E_{\rm kin}^0\simeq-0.849$. This can be used as reference energy.
More specifically one finds
\begin{equation}
  E_{\rm kin}=\sum_{\sigma}
\integral{\epsilon_{\vk}}{}{}\rho_0(\epsilon_{\vk})\epsilon_{\vk}\integral{\omega}{}{}f(\omega)
g_{\vk,\sigma}(\omega),
\end{equation}
where
\begin{equation}
  g_{\vk,\sigma}(\omega)=-\frac1{\pi}\Imag \frac
  1{\sqrt{\zeta_{A,\sigma}(\omega)\zeta_{B,\sigma}(\omega)}-\epsilon_{\vk}}.
\end{equation}

The interaction energies $E_U$, $E_g$ can be calculated from expectation values,
\begin{equation}
  E_U=\frac{U}2\sum_{\alpha}\expval{\hat n_{\alpha,\uparrow}\hat
    n_{\alpha,\downarrow}}{},\;
  E_g=\frac g2\sum_{\alpha}
  \expval{(b_{\alpha}+b_{\alpha}^{\dagger})(\hat n_{\alpha}-1)}{}
\nonumber
\end{equation}
We distinguish between $A$- and $B$-sublattice values, which are equal in the
AFM case, but not for the CO case. There we use the operator identity 
$\hat n_{B,\sigma}=1-\hat n_{A,\sigma}$ (particle hole
transformation), at half filling, which yields
$\expval{\hat n_{B,\uparrow}\hat n_{B,\downarrow}}{}=\expval{\hat
  n_{A,\uparrow}\hat n_{A,\downarrow}}{}+1-\expval{\hat n_{A}}{}$.
The terms for $E_g$ turn out to be equal on $A$ and $B$ sublattice as both
terms will contribute with opposite value.

\section{Phase diagram and static properties of the electrons}
As the electron-phonon coupling in  (\ref{hubholham}) is linear, the bosonic
field can be integrated out in a path integral 
framework, which yields a purely electronic theory with an effective
electron-electron interaction of the form 
\begin{equation}
  U_{\rm eff}(\omega)=U+g^2D^0(\omega),
\label{Ueffom}
\end{equation}
where the free phonon propagator on the real axis is
$D^0(\omega)=2\omega_0/(\omega^2-\omega_0^2)$. 
These are the competing interactions on different
energy scales. There is a limiting case $\omega_0\to \infty$, the
antiadiabatic limit, where $\lambda=2g^2/\omega_0$ is kept fixed.  In this
limit,  $U_{\rm  eff}(\omega)$ becomes independent of $\omega$ and tends to
$U_{\rm   eff}=U-\lambda$, so that the model then becomes equivalent to a
Hubbard model with $U=U_{\rm eff}$. 
Generally, the situation is more complicated.
For large $\omega$ the Coulomb repulsion $U$ is dominant in (\ref{Ueffom}) as
$D^0(\omega)$ goes to zero. However, $\omega_0$ enters as a relevant
energy scale at lower energy and for $|\omega| 
\lesssim \omega_0$ the competition between the bare interactions is most
important. 
The DMFT calculations deal with these competing interactions on different
energy scales, and as a result the ground state phase diagram of the infinite
dimensional HH model at half filling emerges. This phase diagram was already presented
earlier \cite{Bau09pre}. In order to give a comprehensive  account in this
paper, we include some of these results in the following. 

We comment briefly on the notation and energy scales. In electron-phonon physics, specifically
in Migdal-Eliashberg theory, the electron phonon
coupling is usually specified by a dimensionless parameter, often called
$\lambda$, involving the coupling strength and an electronic scale. In contrast,
in this paper $\lambda$ has the dimension of an energy, and it is compared with
the Coulomb repulsion $U$. The dimensionless parameter for electron-phonon 
coupling corresponding to the usual convention would be
$\bar\lambda=\rho_0(0)\lambda$, where $\rho_0(0)=2/(D\pi)$.
For physical optical phonons the scale $\omega_0$ is expected to be roughly
$\omega_0\sim0.01W-0.2W$, where $W$ is the electronic bandwidth.
The purpose of the paper is to characterize the electron-phonon system quite
generally, with a tunable parameter $\omega_0$, which for many results we
chose as $\omega_0=0.15 W$ neither too close to the adiabatic nor to the
antiadiabatic limiting cases for illustration of intermediate behavior. It
should be noted that the value is large for most realistic electron-phonon
systems with the possible exception of the alkali doped fullerides
\cite{gunnarsson}. For the former cases smaller values such as $\omega_0=0.05
W$ could serve as a better guideline. 

\subsection{Phase diagram}
\begin{figure}[!htbp]
\centering
\includegraphics[width=0.45\textwidth]{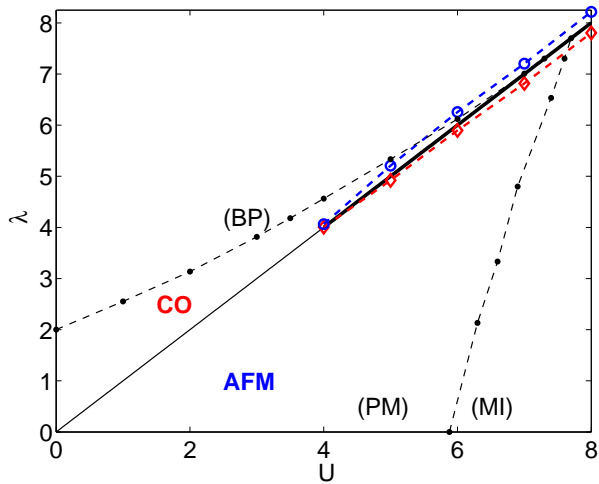}
\caption{(Color online) The phase diagram of antiferromagnetic (AFM) and
  charge order (CO) in the $U$-$\lambda$-plane. The thin black
  line $\lambda \simeq U$ gives a continuous transition and the thick line a
  discontinuous one. A non-zero order parameter
  was found above dashed line with diamonds for CO and below the dashed line
  with circles for AFM order. The  dashed lines with points gives the
  transition for phases with no long range order, a paramagnetic metallic
  (PM), bipolaronic (BP) and Mott insulator (MI).}
\label{hfphasediag}
\end{figure}

We first present the complete phase diagram as shown in
Fig. \ref{hfphasediag}. The overall energy scale is set by the bandwidth
$W=4t=4$, and the phonon frequency $\omega_0=0.6 t$ was chosen.
Many results such as the global
phase diagram are similar for different choices of $\omega_0$, but we will
also point out the differences appearing for other values of $\omega_0$.
The limiting cases of the phase  diagram can be  understood on a qualitative
level. The $U$-axis corresponds to the pure repulsive
Hubbard model which is known to be
AFM ordered at weak coupling \cite{Don91} for a bipartite lattice, and this
order is smoothly connected to the strong coupling Heisenberg AFM
\cite{ZPB02,BH07c}. Also along the $\lambda$-axis, the corresponding  pure
Holstein model has a charge ordered ground state for 
$g>0$ and $\omega_0>0$, such that the limits of weak and strong
coupling are smoothly connected \cite{FJS93}. 
For finite $U$ and $g$ we find that the transition line is close to the 
line $\lambda= U$ with a small tendency towards $\lambda >U$.   
For larger values of the interactions we have included a dashed line
($\diamond$) above which our DMFT-NRG calculations find solutions with 
finite $\Phi_{\rm co}$ and another one ($\circ$) below which $\Phi_{\rm afm}$  
is finite. 

\noindent
For  weaker coupling ($U<3$) the order parameters become very
small  close to the line $U_{\rm eff}= 0$.
There are strong indications that the transition proceeds both directly and continuously 
from one type of  ordered to the other ordered state, such that only at the
critical point both order parameters vanish. For instance, we find that all
the  relevant response quantities behave continuously.
A direct order to order transition is found in the antiadiabatic limit, where
for any finite $U_{\rm eff}>0$ the system is AFM ordered \cite{Don91} and for
$U_{\rm eff}<0$ in the CO or SC state \cite{MRR90}. We have 
therefore a continuous transition from an ordered to an ordered state with
vanishing order parameters at the transition. There
is no reason to expect that the transition becomes discontinuous immediately when
$\omega_0$ is decreased from infinity to a finite value $\omega_0\gg W$.
In addition, in the DMFT-NRG calculations we find that near the transitio, the
smaller $\omega_0$ is, the larger the order parametern becomes, as detailed
later in Fig. \ref{phi_lamdepU3varwo1}. Thus we conclude that the
direct continuous transition scenario persists for weak coupling and finite 
$\omega_0$. 
Mean field calculations (see Appendix \ref{mfapp}) in the adiabatic limit support 
the picture of a direct transition  from one ordered state to the
other, however, the transition is always discontinuous then. Also the local
effective quasiparticle interaction $U^r$ (which will be discussed 
more fully  later) is observed to change sign
at $U_{\rm eff}= 0$,   which is consistent with a change of ground state
there.  

For larger couplings we have a parameter regime where we can find finite
$\Phi_{\rm afm}$ and $\Phi_{\rm co}$. In this regime the transition from the
AFM to the CO turns out to be discontinuous and one can identify a point on
the transition where the nature of transition changes. 
The calculation of the total ground state energy (see inset of
Fig. \ref{phi_lamdepU5}) shows that also here the transition occurs
approximately at $U_{\rm eff}\simeq 0$. The behavior of the model along the
line $U=\lambda$ has been studied in Ref. \cite{BH10a}.

The HH model in one dimension has been studied in great detail with efficient
numerical methods \cite{CH05,HC07,FHJ08}.
One finds a Mott insulator with strong antiferromagnetic
correlations, but no long range order,  when $U_{\rm eff}>0$, and a Peierls
charge density wave (CDW) insulator for $U_{\rm eff}<0$. There is, however, a metallic region with
finite spin gap, but no charge gap in between these two phases. The transition
line to the CDW state appears for values of $\lambda$ a bit larger than $U$,
similar to what is observed in our calculations but more pronounced.
For larger $U$ this intermediate region shrinks until we get a direct first
order Mott-Peierls transition. 
A major difference with the high dimensional results is the real symmetry
breaking in our case as well as the existence of the intermediate region, for
which we find no indication here. 
Nevertheless the mentioned similarities and also recent results in $d=2$
(adiabatic limit)\cite{KB08} suggest that the general features of the phase diagram in
Fig. \ref{hfphasediag} might be quite general, largely independent of
dimensionality.

In Fig. \ref{hfphasediag} we have also included the phase boundaries of the
HH model when no long range order is allowed for. We see that only for large
coupling do the phase boundaries merge, while for smaller couplings other scales
are important. For the case $\omega_0=0.2$ this has been analyzed earlier
\cite{JPHLC04,KMH04,KMOH04}. The Mott transition, as obtained on increasing $U$
for fixed $g$ is only little affected by the additional electron-phonon
coupling, which is manifested in a shift of the critical $U_c$ for the
transition. The metal-bipolaron transition, observed when increasing $g$ for
fixed $U$, is of second order for smaller interactions and becomes of first
order for large interactions, and thus similarities with the ground state
behavior are found. Note that the CO state and the bipolaronic state are
different, as in the latter no symmetry is broken and the occupation
expectation value is always 1, whereas in the CO state $n_A\neq n_B$. For
details we refer to Refs. \onlinecite{JPHLC04,KMH04,KMOH04}.

\subsection{Order parameter as a function of $\lambda$}

\begin{figure}[!htbp]
\centering
\includegraphics[width=0.45\textwidth]{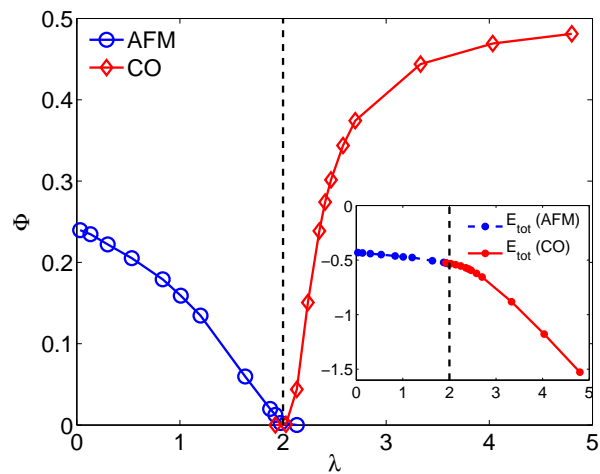}
\caption{(Color online) The expectation values $\Phi$ for $U=2$ as a function
  of $\lambda$. The inset shows the  total energy.}       
\label{phi_lamdepU2}
\end{figure}

We consider here the way the two types of order parameter
change as a function of $\lambda$ for fixed values of $U$ both in the weak
coupling and strong coupling regime.
A weaker coupling  case with $U=2$ is shown in
Fig. \ref{phi_lamdepU2}.

\begin{figure}[!thpb]
\centering
\includegraphics[width=0.45\textwidth]{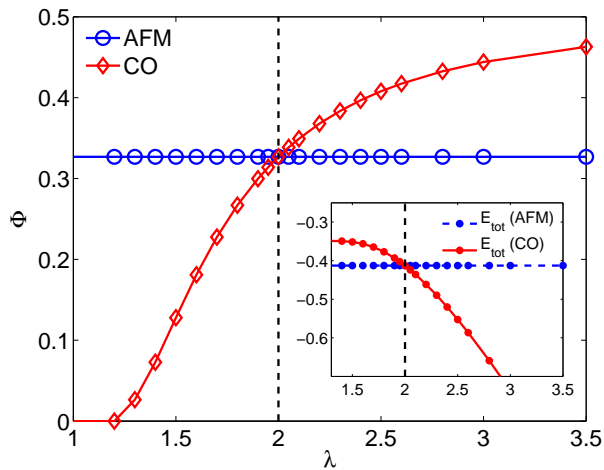}
\caption{(Color online) Mean field result for the order parameter and total
  energy shown in the inset.}       
\label{mfphi_lamdep_U2}
\end{figure}

\noindent
We can see that the AFM order is decreased when the
electron phonon-coupling is increased, as the repulsion is reduced.
Near $\lambda=U$ the ordering scale is very small
($<10^{-3}$) and cannot be resolved in our DMFT-NRG calculations. For
$\lambda>U$ the $\Phi_{\rm co}$ shows a steep rise with $\lambda$.
For this  weak coupling case we can study the limit $\omega_0\to 0$ and compare
with the corresponding  static mean field 
theory (details see appendix \ref{mfapp}). For $U=2$
the numerical results are shown in Fig. \ref{mfphi_lamdep_U2}. 

\noindent
The solutions are for situations where the order is exclusive, i.e. only one of
the order parameters is nonzero.
In the mean field calculation the AFM order parameter is larger
than in the DMFT case, and when $\lambda$ is increased is seen not to be affected by the
electron phonon coupling as long as $\Phi_{\rm co}=0$. In contrast the CO
order state $\Phi_{\rm   co}$ feels the $U$-term, but increases with
$\lambda$. For $U=\lambda$ the mean field equations give order parameters that
coincide, and once $\Phi_{\rm 
  co}$ exceeds $\Phi_{\rm afm}$ the charge ordered state possesses the lowest
energy as can be seen from equation (\ref{Emfsimp}) and the inset in
Fig. \ref{mfphi_lamdep_U2} . We can infer from this 
weak coupling result that there is a direct, discontinuous 
transition from  an ordered to an ordered state at $U=\lambda$ for
$\omega_0\to 0$.
As seen in Fig. \ref{phi_lamdepU2} the behavior is strongly modified when quantum
fluctuations are included and $\omega_0$ is well finite, as the order
parameters are influenced much more by the presence of the competing
interaction.  It is numerically not possible to study the limit $\omega_0\to
0$ within our DMFT approach due to the increase in the bosonic Hilbert
space. The analysis for different values of $\omega_0$ is however conform with
the trends discussed here (see Fig. \ref{phi_lamdepU3varwo1}). 

\begin{figure}[!htbp]
\centering
\includegraphics[width=0.45\textwidth]{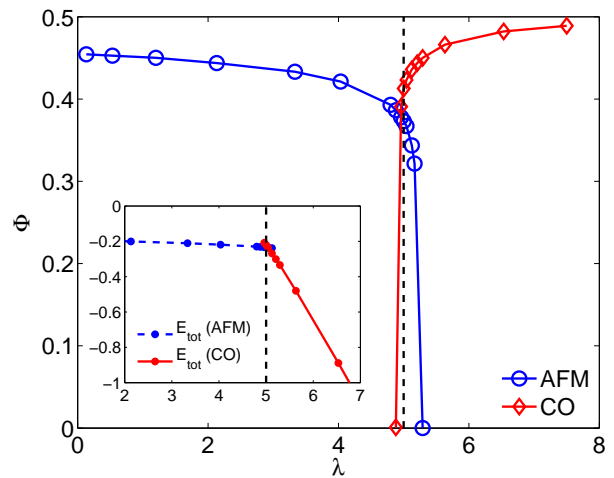}
\caption{(Color online) The expectation values $\Phi$ for $U=5$ as a function
  of $\lambda$.}       
\label{phi_lamdepU5}
\end{figure}

For larger couplings we saw that the two dashed lines in
Fig. \ref{hfphasediag} cross, which means that we have a parameter regime
where we find finite $\Phi_{\rm afm}$ and $\Phi_{\rm co}$ . An example for
this behavior is shown in Fig. \ref{phi_lamdepU5} for $U=5$. 

\noindent
The transition here is seen to be rather sharp.
The calculation of the total ground state energy (see inset) shows that the
transition occurs approximately  at $U_{\rm eff}= 0$, in fact it occurs for
small negative $U_{\rm eff}$, i.e., on the $\lambda>U$ side. A number of quantities
such as the double occupancy $\expval{\hat n_{\uparrow}\hat n_{\downarrow}}{}$
(see section \ref{docc}) show discontinuities at the transition. The total
energy (see inset of Fig. \ref{phi_lamdepU5}) is continuous function of
$\lambda$. It shows, however, a kink at the transition, such that a first
derivatives will be discontinuous. 

\subsection{Double occupancy}
\label{docc}
\begin{figure}[!htbp]
\centering
\includegraphics[width=0.45\textwidth]{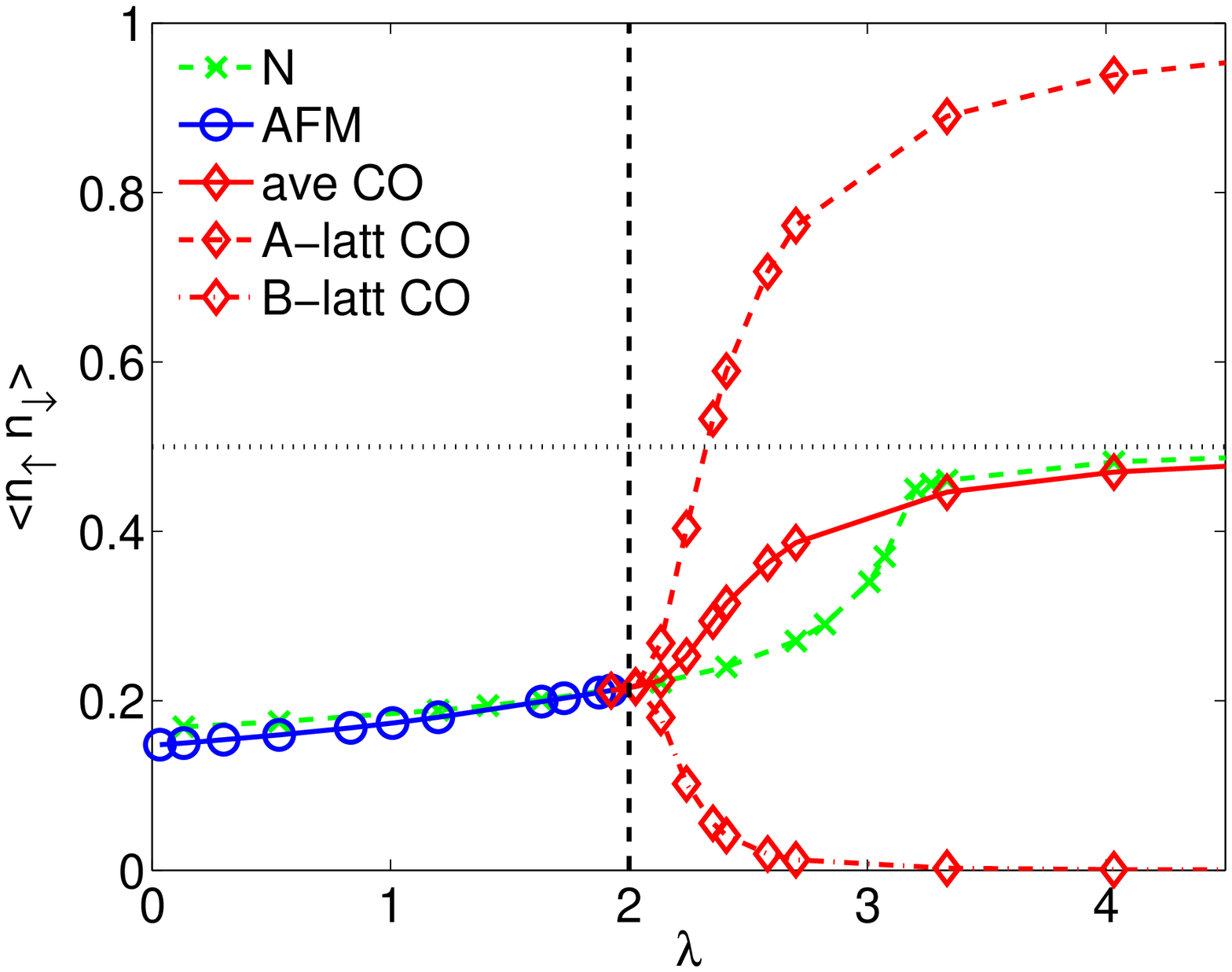}
\includegraphics[width=0.45\textwidth]{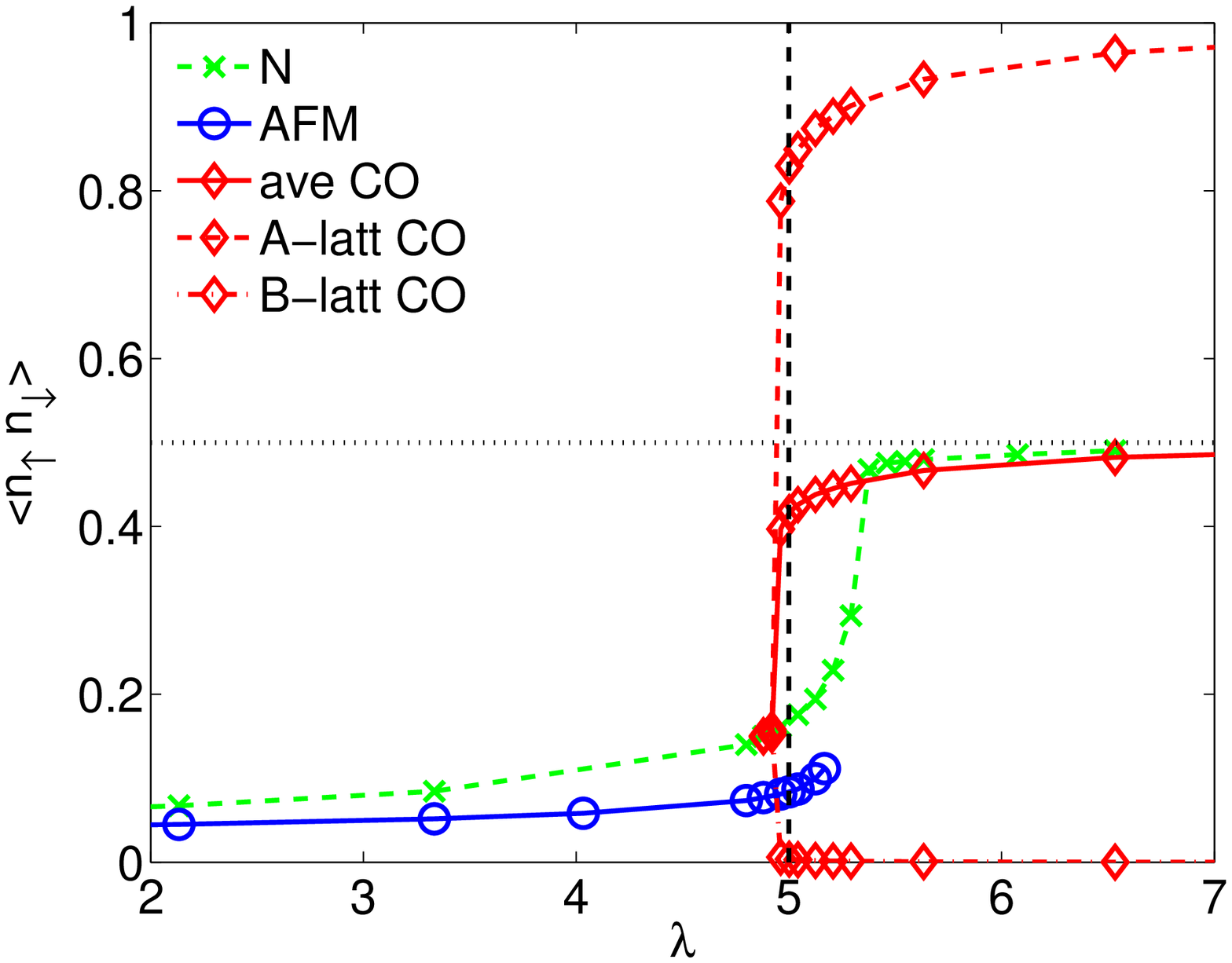}
\caption{(Color online) The expectation values
  $\expval{n_{\uparrow}n_{\downarrow}}{}$ for $U=2$ (upper panel) and $U=5$
  (lower panel) as a function of $\lambda$ for the N, AFM and CO state.}       
\label{docc_lamdepU25}
\end{figure}

A characteristic quantity for the electronic part of the system is the 
expectation value for the local double occupancy.
This is a homogeneous quantity in the normal (N) and AFM phase. In the charge ordered
phase it differs for $A$- and $B$-sublattice, and $\expval{\hat
  n_{A,\uparrow}\hat n_{A,\downarrow}}{}$ and $\expval{\hat n_{B,\uparrow}\hat n_{B,\downarrow}}{}$
 are given as detailed above when discussing
the energy due to the Hubbard interaction term. We can compare the quantities by
taking the average over the two sublattices,
$\expval{\hat n_{\uparrow}\hat n_{\downarrow}}{}=\sum_{\alpha}\expval{\hat
  n_{\alpha,\uparrow}\hat n_{\alpha,\downarrow}}{}/2$. In Fig. \ref{docc_lamdepU25}
we show the results for $U=2,5$ as a function of $\lambda$. We 
have included the N, AFM and CO state, and for the latter, the averaged as well
as the sublattice quantities.

\noindent
For the continuous transition at $U=2$ we find a small nearly linear increase
with $\lambda$ for the N and AFM state. The values for N and AFM
state are very similar with the double occupancy being larger in the normal phase,
where the charge carriers are more mobile than in the ordered phase. This is
different from the strong coupling case, where the N state is in the Mott
phase, and  the double occupancy is lower than in the AFM case. This is due to
the super-exchange mechanism which leads to a kinetic energy gain and a
slightly higher mobility and also double occupancy. When 
$\lambda>U$ and charge order sets in, the double occupancy shows a steep rise on
the $A$-sublattice and decrease on the $B$-sublattice. As can be seen in
the top panel of Fig. \ref{docc_lamdepU25}, the $\lambda$-dependence is
continuous for $U=2$. At 
approximately $\lambda=3$ the N state metal-bipolaron transition occurs, where
$\expval{\hat n_{\uparrow}\hat n_{\downarrow}}{}$ in the normal state increases rapidly
(but also continuously). $\expval{\hat n_{\uparrow}\hat n_{\downarrow}}{}$ is
then larger than in the CO state which can be understood by thinking about the
fact that the CO state wins energetically against the BP state through kinetic
energy due to pair hopping (see Sec. \ref{difenergies}).  

The overall behavior in the lower panel of Fig. \ref{docc_lamdepU25} for
$U=5$ is similar. $\expval{\hat n_{\uparrow}\hat n_{\downarrow}}{}$ increases
slightly with $\lambda$ for a certain range and more rapidly near the
transition. There, we can clearly see the discontinuous behavior for
$\lambda\simeq 5$ and $\expval{\hat n_{\uparrow}\hat n_{\downarrow}}{}$ jumps
from the value $0.09$ in the AFM 
state to $0.42$ in the CO state. The BP state again has a larger double
occupancy than the average of the CO state. 

\subsection{Order parameter as a function of $U$}
\begin{figure}[!htbp]
\centering
\includegraphics[width=0.45\textwidth]{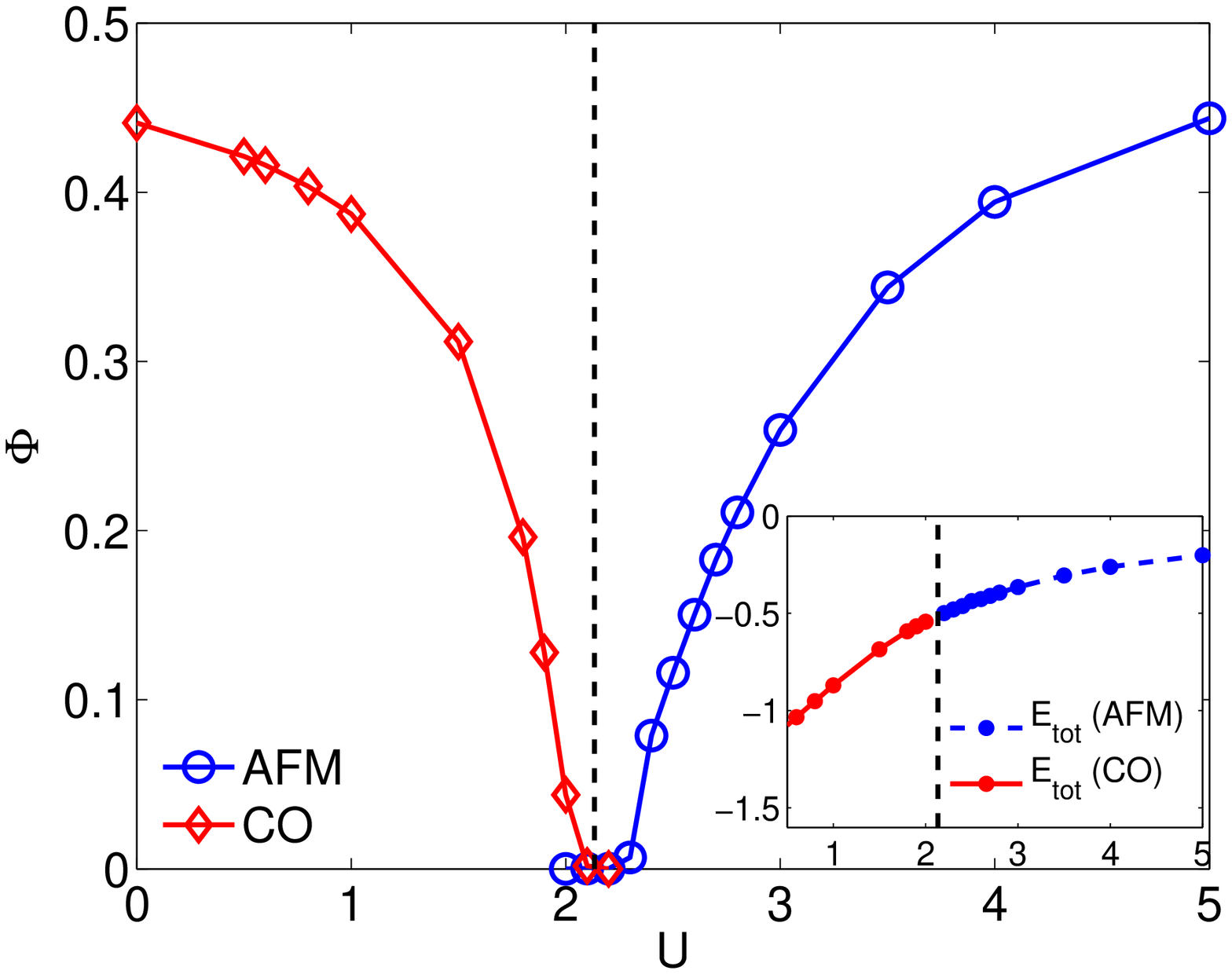}
\includegraphics[width=0.45\textwidth]{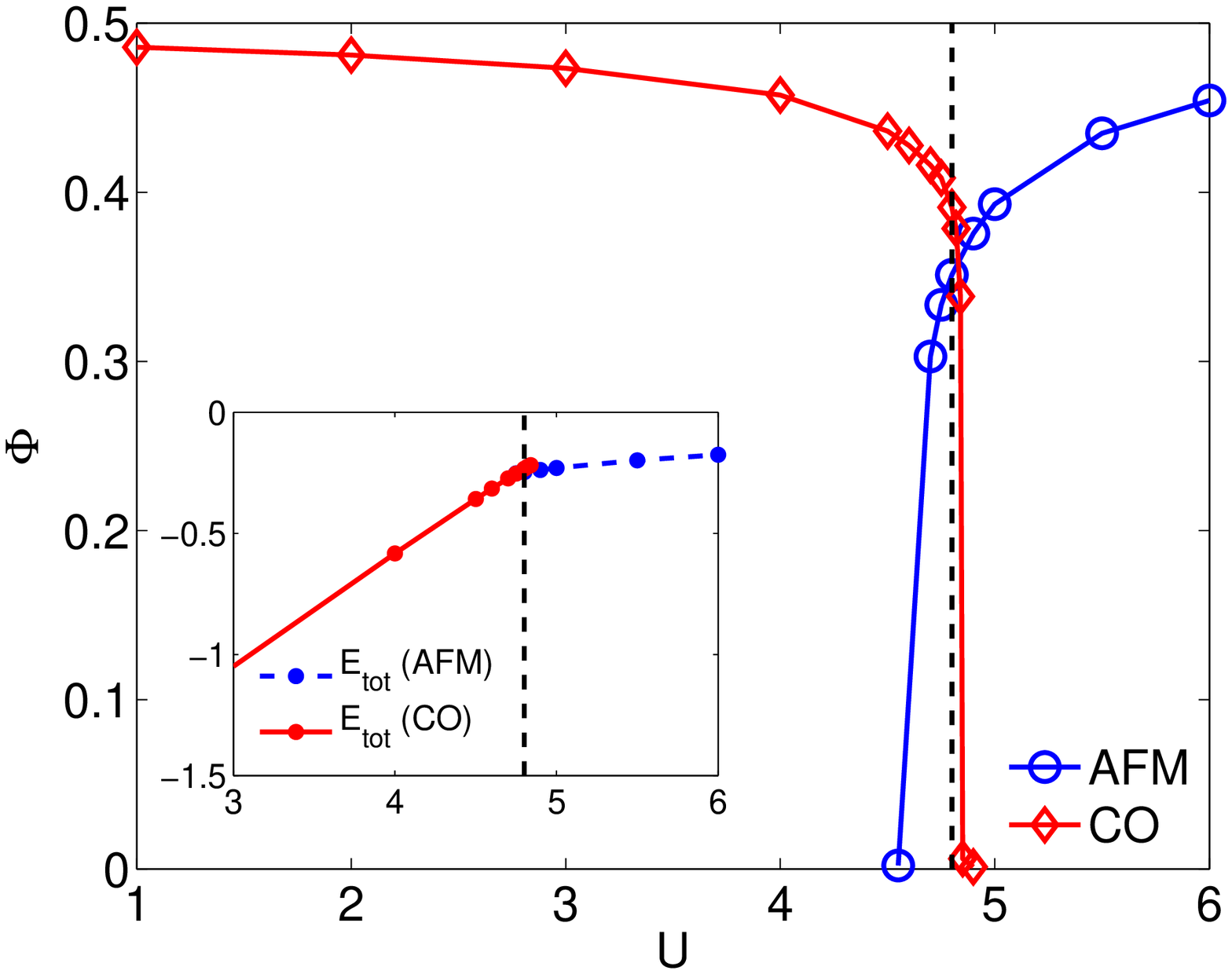}
\caption{(Color online) The expectation values $\Phi$ for $\lambda\simeq 2.13$
  (top) and $\Phi$ for $\lambda\simeq 4.8$ (bottom) as a function of $U$. The
  insets show the total ground state energy.}        
\label{phi_Udep_g0.8g1.2}
\end{figure}

So far we have analyzed the transition as a function of $\lambda$. Similarly,
we can look at the properties of the model as a function of the on-site repulsion 
$U$. We focus on the order parameters and look at the cases of a fixed
$\lambda\simeq 2.13$ ($g=0.8$) and $\lambda\simeq 4.8$ ($g=1.2$) shown in
Fig. \ref{phi_Udep_g0.8g1.2}. 

\noindent
In both cases for small $U$ the CO state dominates, but on increasing $U$,
$\Phi_{\rm co}$ is driven to zero and $\Phi_{\rm afm}$ becomes finite.
Similar to the weaker coupling case for fixed $U$ for $\lambda\simeq 2.13$,
the order parameters approach zero at the transition. We also have continuous transition here, which is
visible in the total energy plotted as an inset.

In contrast for larger $\lambda\simeq 4.8$  (lower panel in
Fig. \ref{phi_Udep_g0.8g1.2}) the picture is as for the larger coupling cases
above, where the transition occurs with finite order parameters, which change
discontinuously.   
Also here, as seen in the inset, the total ground state energy displays a 
kink at the transition. 
Summarizing, we can say that the transition occurs in
similar fashion as a function of $U$ or $\lambda$.  Depending on the
magnitude of the coupling constants continuous or discontinuous transitions can be
observed.

\subsection{Dependence of the order parameter on $\omega_0$}
\begin{figure}[!htbp]
\centering
\includegraphics[width=0.45\textwidth]{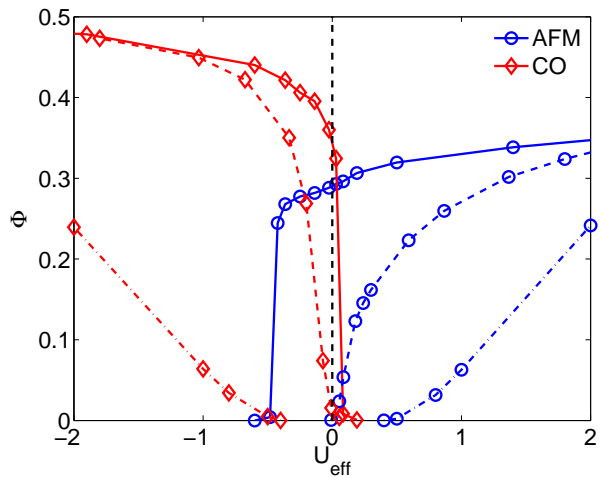}
\caption{(Color online) The expectation values $\Phi$ for a range of effective
  $U_{\rm eff}=-2,2$ for fixed $U=3$ varying $g$. We show the results for  $\omega_0=0.2$ (full line),
  $\omega_0=0.6$ (dashed line), and $\omega_0\to\infty$ (dot-dashed line).}       
\label{phi_lamdepU3varwo1}
\end{figure}

So far we have found that the transition from the CO to the AFM state occurs close to
$U_{\rm eff}=0$ independently of the other parameters. For fixed $\omega_0$,
when studying the order parameters or double occupancy, one could see quite a
different dependence on $\lambda$ or $U$ near the transition with a continuous
and discontinuous behavior.  In the following we study how the
order parameters depend on $U_{\rm eff}$ near the transition for different
values of $\omega_0$.  
We illustrate this in Fig. \ref{phi_lamdepU3varwo1}, where we plot the
respective order parameters for different $\omega_0$, including the case
$\omega_0\to\infty$, which is given by the pure Hubbard model with local
interaction $U_{\rm eff}$. We have held $U=3$ fixed and varied $g$ to obtain
the desired values.

\noindent
As one can see clearly the ``sharpness''  of the transition increases
when $\omega_0$ is decreased. 
For $\omega_0\to\infty$ the order parameters
approach zero in a similar exponential form as in mean field theory \cite{BHD09}. The form
of the transition is then symmetric with respect to $U_{\rm eff}=0$. For
finite $\omega_0$ when tuning $U_{\rm eff}$ with $g$ the transition will be
asymmetric and the order parameter is decreased much less when the effective
interaction is close to 0, as seen most 
pronounced for the case $\omega_0=0.2$ (full line). In the
adiabatic limit, $\omega_0\to 0$, we always expect a discontinuous transition
at zero temperature. For finite $\omega_0$ we have the competition of the
Hubbard repulsion and phonon-induced attraction. The effect of latter enters
at lower energies for smaller values of $\omega_0$. This might explain why the
AFM order is more stable then. To establish CO as the ground state, it 
seems to be mainly necessary that $U \lesssim \lambda$, and the large $U$ at
higher energies does not spoil this. Retardation effects seem to play hardly any role
at half filling for these static orders. 

We can conclude from this section that with generality the AFM-CO quantum
phase transition occurs approximately when electron interaction parameter $U$
and phonon attraction $\lambda$ are equal. The behavior near the transition,
e.g. the order parameter depends, however, very much on the the interaction
strength as well as the phonon frequencies. Small phonon frequencies and large
interactions lead to discontinuous behavior, whereas for large phonon
frequencies the competing interactions lead to more cancellations, reduced
order and much evidence for continuous transitions. Then there exists a point on the
transition line $\lambda_{\rm tc}\simeq U$, separating continuous and
discontinuous transitions. We find that the value $\lambda_{\rm tc}$ increases
with $\omega_0$. In the limiting case of $\omega_0\to \infty $, there are only
continuous transitions such that $\lambda_{\rm tc}=\infty$, whereas we expect
that for $\omega_0\to 0$ there are only discontinuous transitions and
$\lambda_{\rm tc}=0$. It would be of interest to explore how 
$\lambda_{\rm tc}$ varies for finite $\omega_0$ as a function of temperature. As temperature
tends to increase fluctuations and decrease order, the naive expectation
would be that $\lambda_{\rm tc}$ increases with temperature.

\section{Different contributions to the total energy}
\label{difenergies}
We have used the total ground state energy $E_{\rm tot}$  to decide whether the AFM or CO
state is the ground state. In this section we give details of the different
contributions to  $E_{\rm tot}$ in 
(\ref{Etot}), and their  dependence on the electron-phonon coupling for fixed
$U$. Let us first remark generally
on the energy of the ordered state in comparison with the normal (N) state.
In the half filled pure Hubbard model at weak coupling the AFM
state has lower potential energy than the N state, but higher kinetic
energy. At strong coupling the AFM has lower kinetic energy than the N 
state (exchange term), but higher potential energy than the N state (Mott
insulator). In the pure Holstein model the CO state has a lower potential energy at weak
and intermediate coupling and higher kinetic energy due to localization. At
strong coupling when the N state is insulating (bipolaronic) the energy
is lowered in the CO state due to lower kinetic energy (pair hopping).

In the situation with finite $U$ and $g$, we have a competition between the different terms. The AFM
state will usually have smaller $E_U$, since double occupancy is
lower, whereas the CO state possesses larger $E_U$ but contributions from $E_g$ lower the
energy. However, we also have to take into account the contribution $E_{\rm
  ph}$, which is larger in the CO state.  

\begin{figure}[!htbp]
\centering
\includegraphics[width=0.23\textwidth]{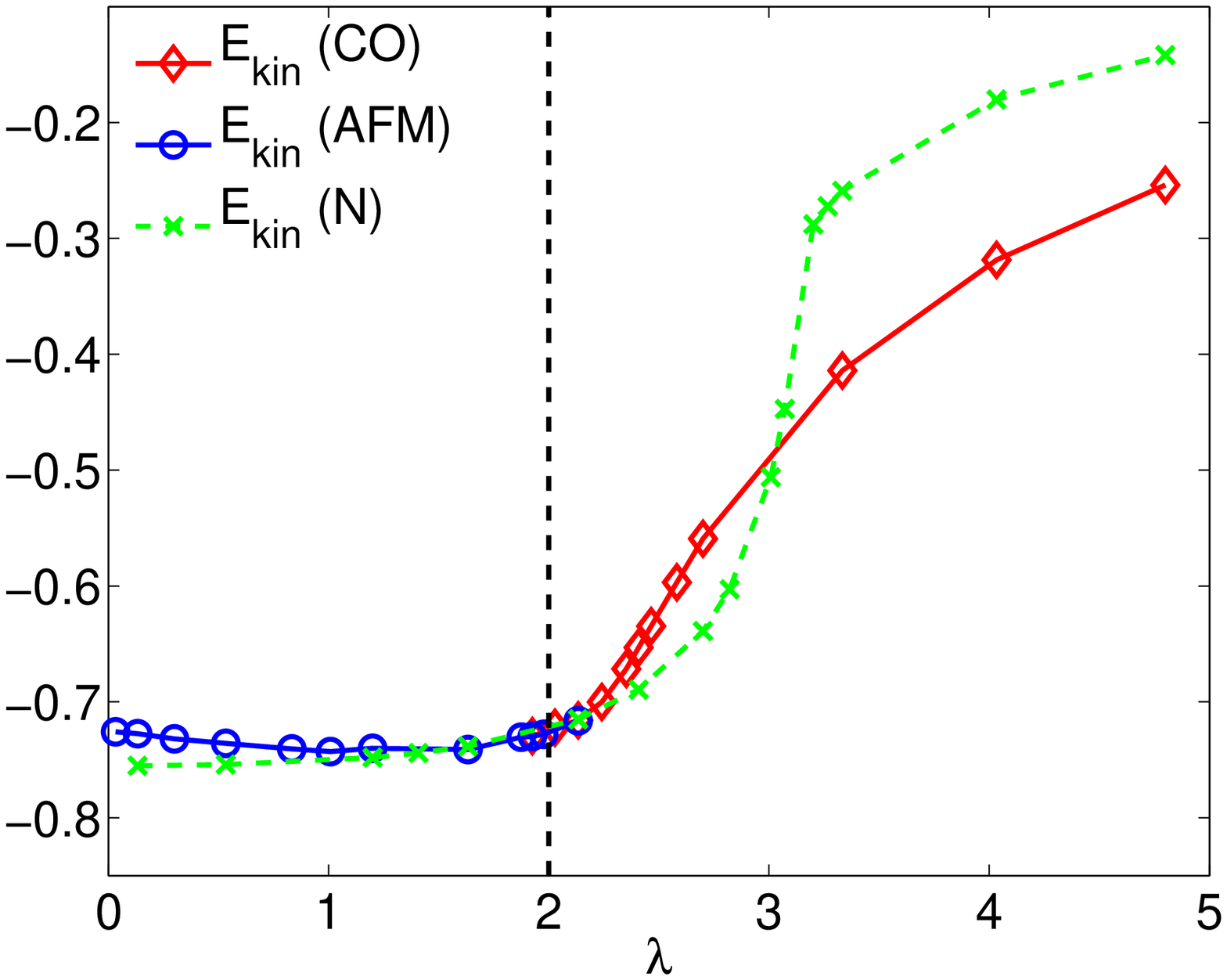}
\includegraphics[width=0.23\textwidth]{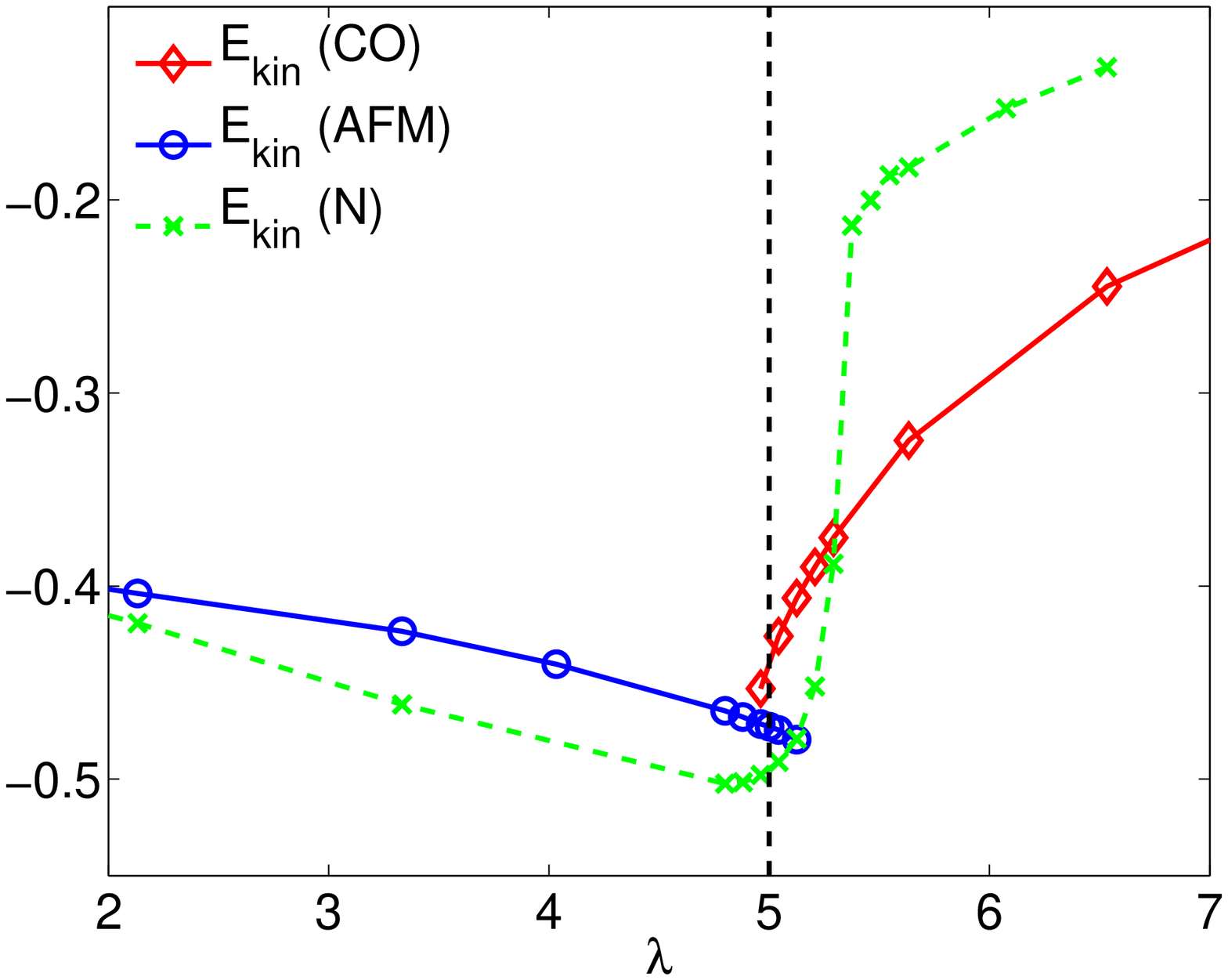}
\includegraphics[width=0.23\textwidth]{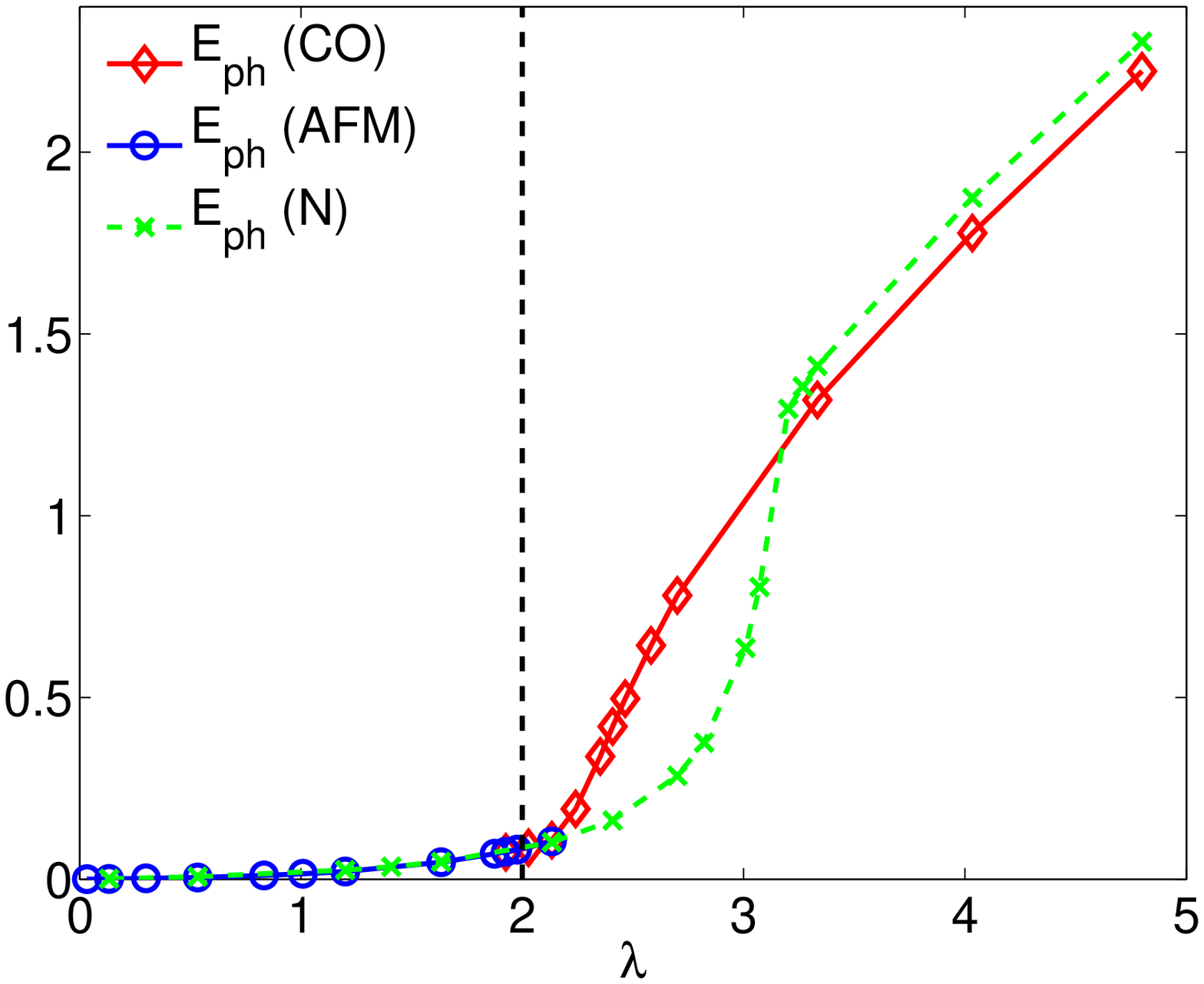}
\includegraphics[width=0.23\textwidth]{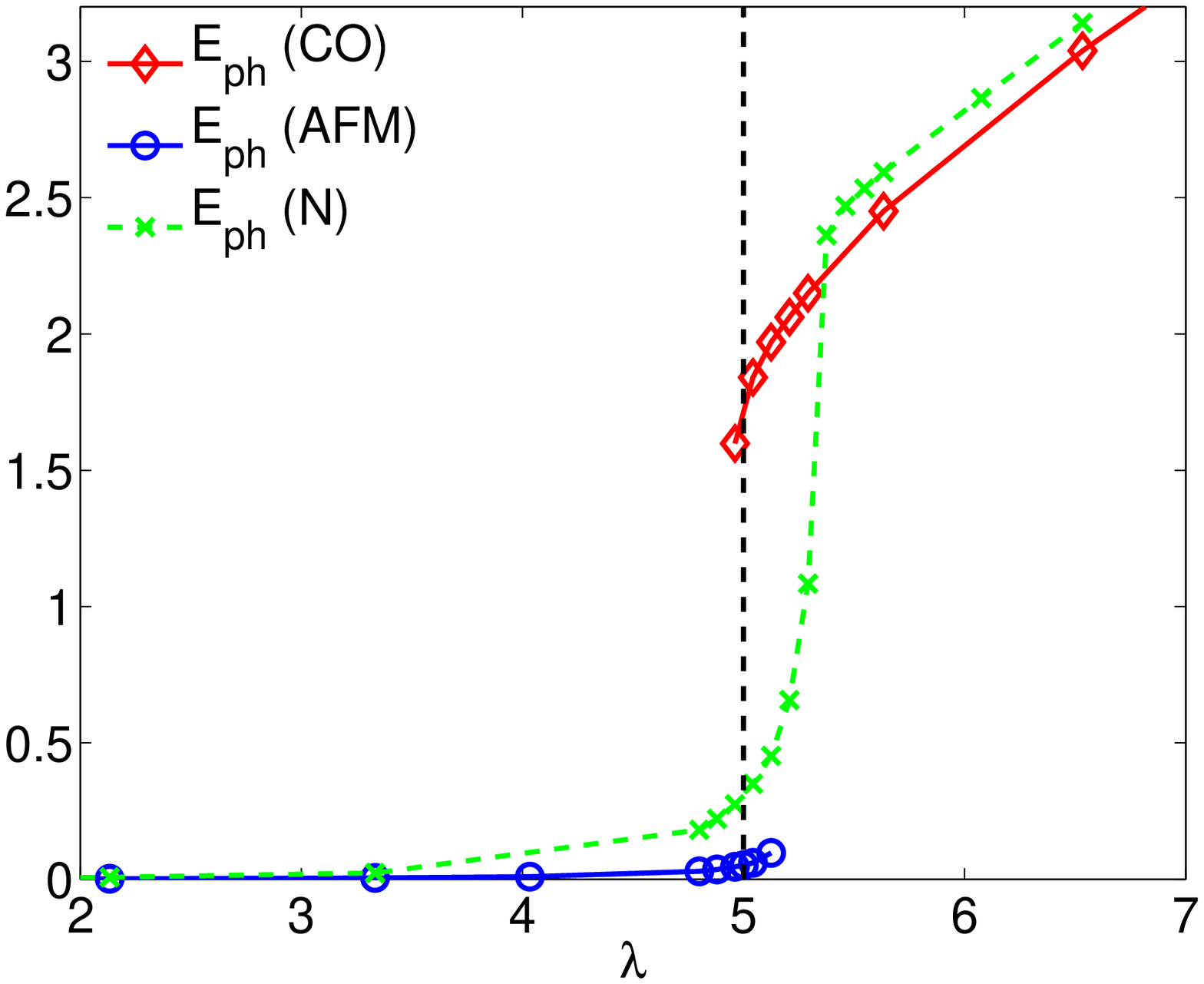}
\caption{(Color online) Comparison of the kinetic energy $E_{\rm kin}$ and phonon energy
  $E_{\rm ph}$ for $U=2$ (left) and $U=5$ (right) as a function of $\lambda$ in
  CO, AFM and N state.}       
\label{engkinph_lamdepU25}
\end{figure}

In Fig. \ref{engkinph_lamdepU25} we see the behavior of kinetic energy for
the electrons $E_{\rm kin}$ and oscillator energy of the phonons $E_{\rm
  ph}$, where the energy $\omega_0/2$ for zero point motion was omitted. 

\begin{figure}[!htbp]
\centering
\includegraphics[width=0.23\textwidth]{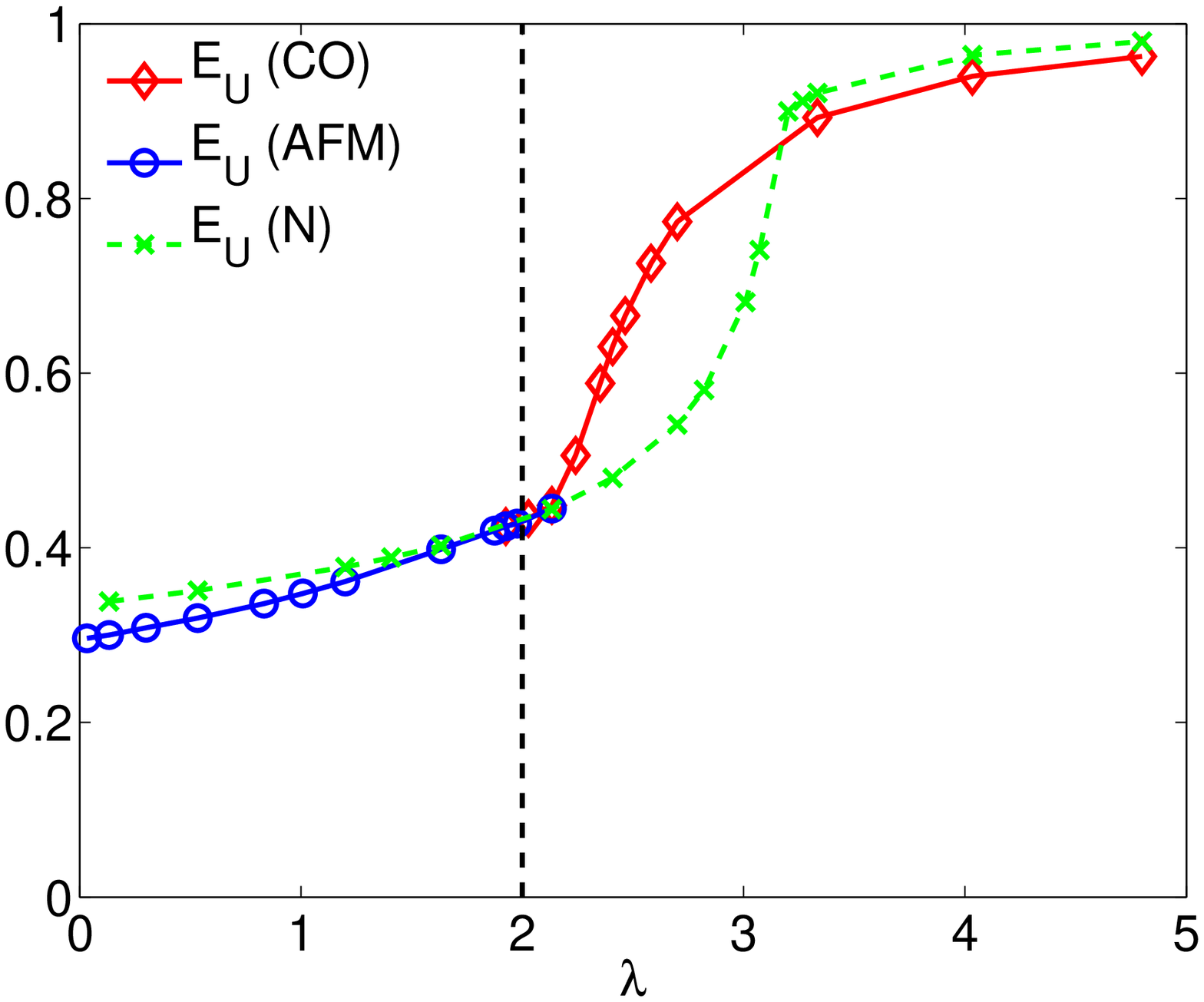}
\includegraphics[width=0.23\textwidth]{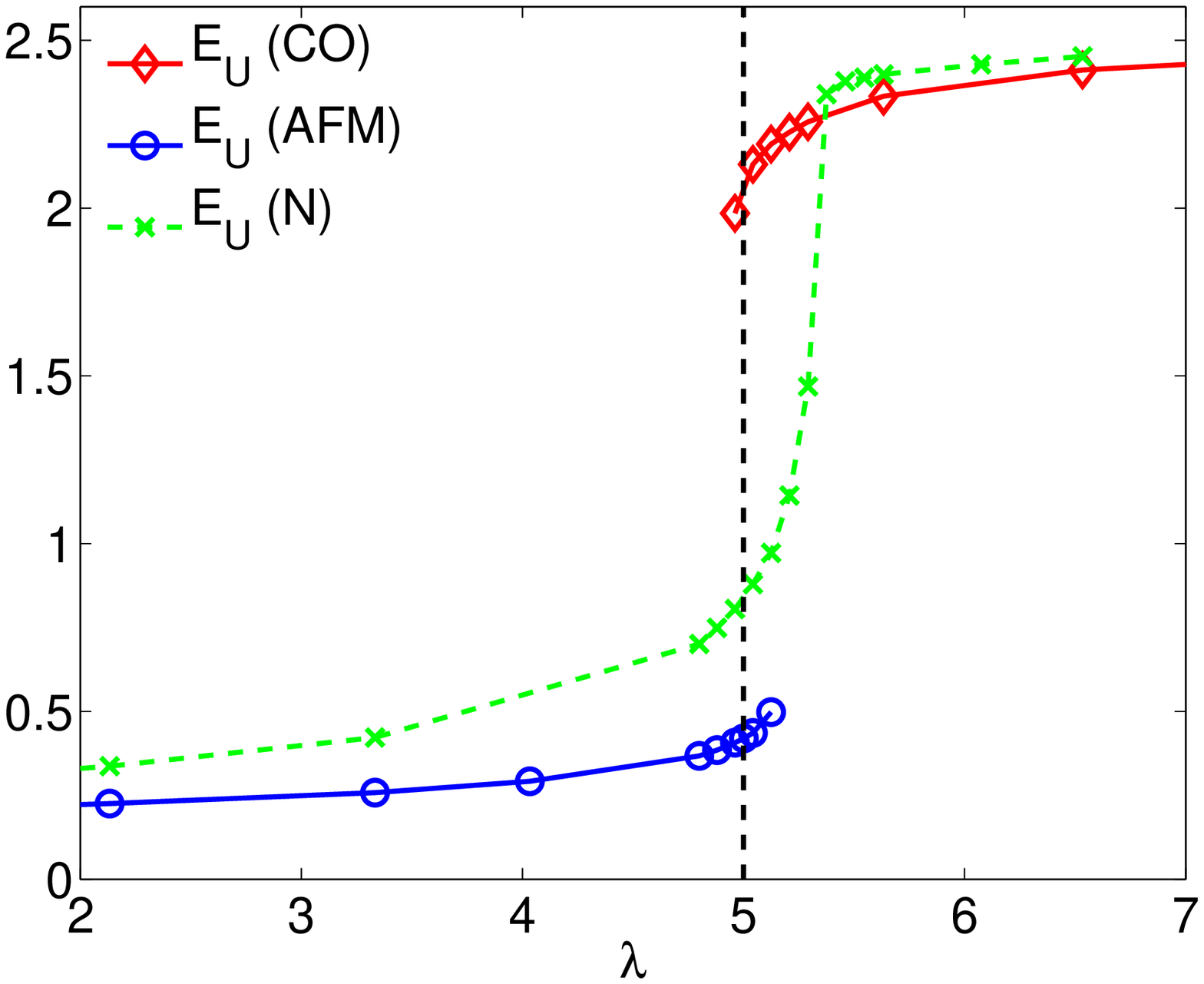}
\includegraphics[width=0.23\textwidth]{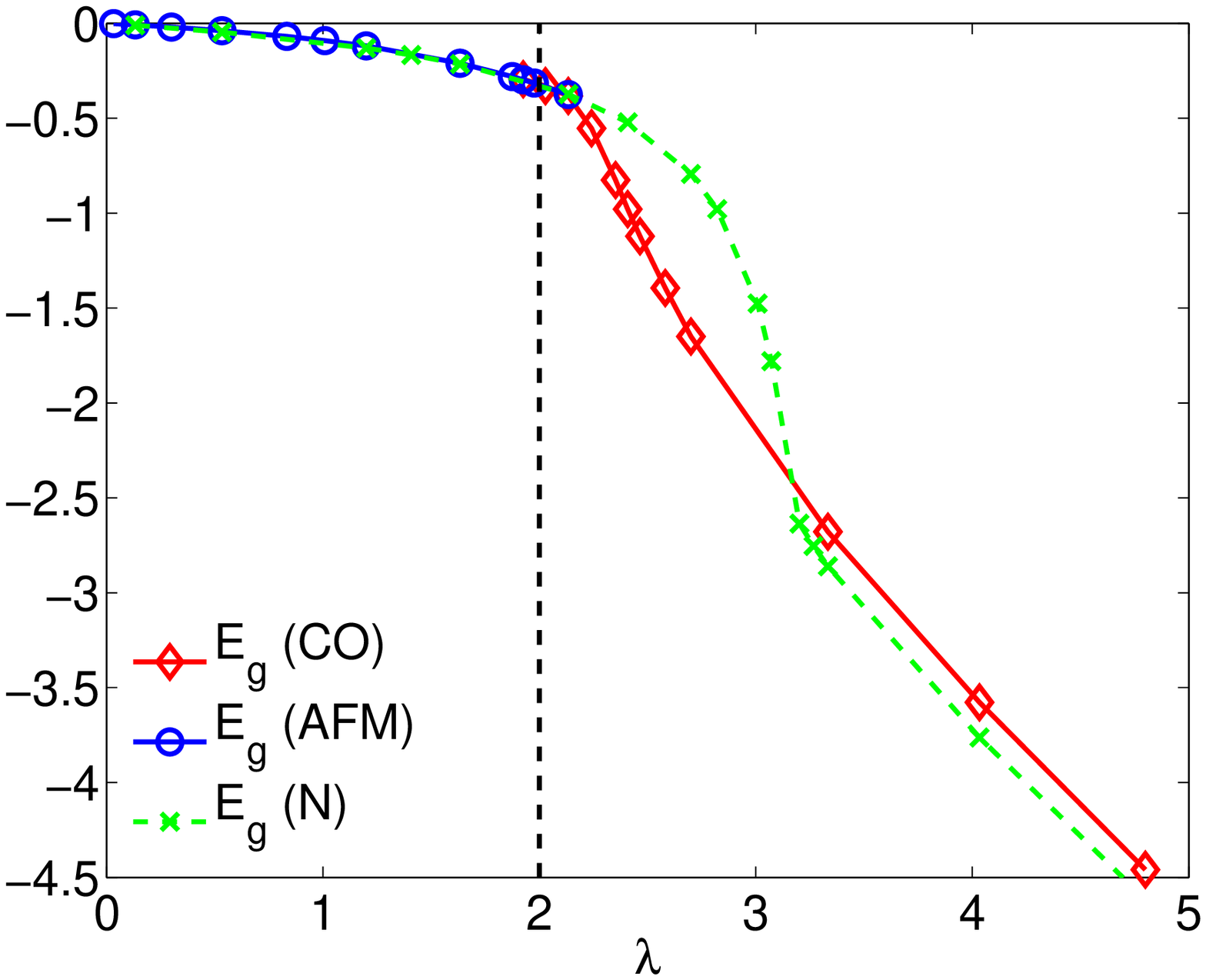}
\includegraphics[width=0.23\textwidth]{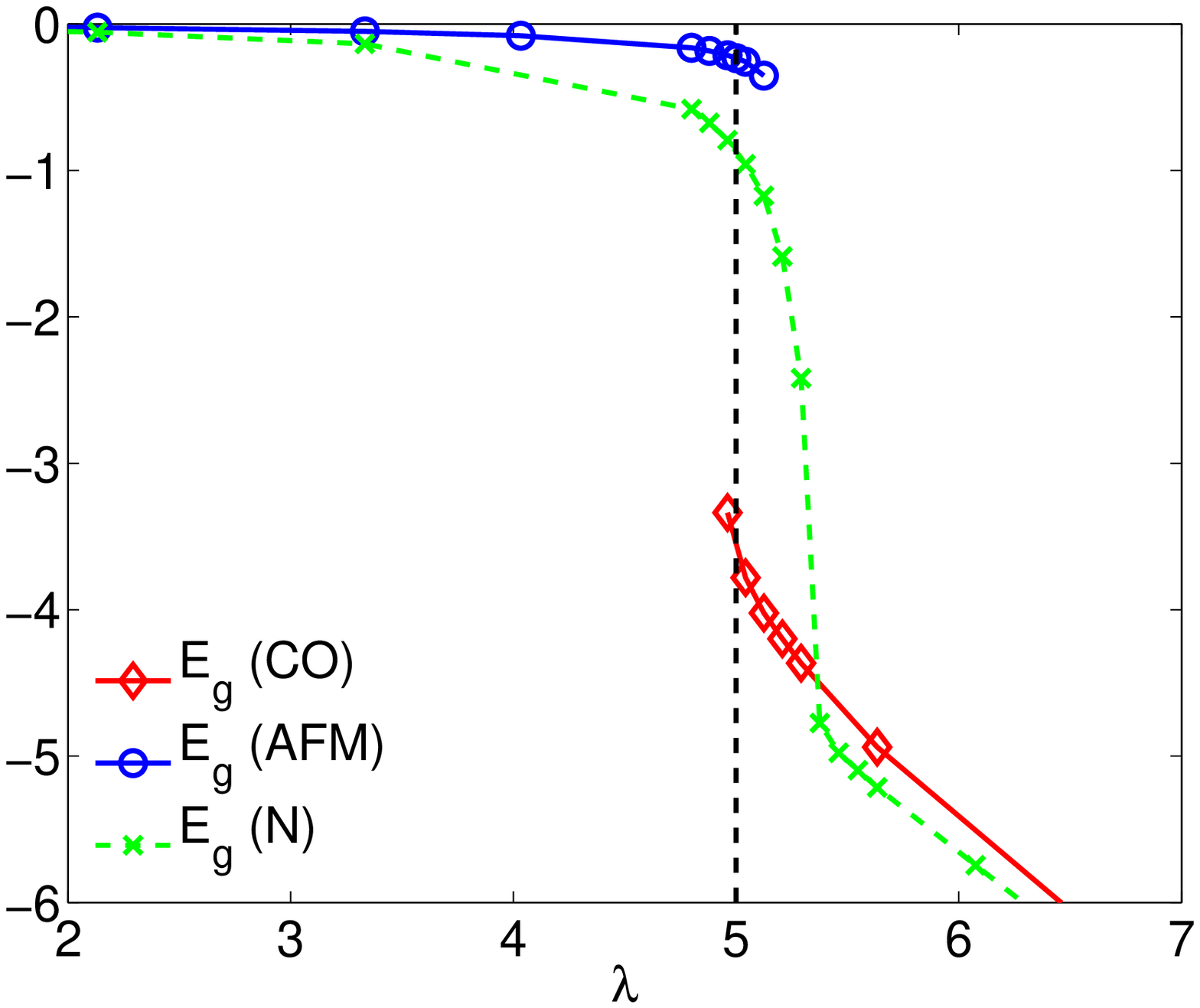}
\caption{(Color online) Comparison of the interaction energies $E_{U}$ and
  $E_{g}$ for $U=2$ (left) and $U=5$ (right) as a function of $\lambda$ in CO,
  AFM and N state.}        
\label{engUg_lamdepU25}
\end{figure}

\noindent
We show the quantities for the values $U=2$ (left) and $U=5$ (right) 
as a function of $\lambda$ in CO, AFM and N state. As a reference energy
recall that for the free system $E_{\rm kin}^0\simeq-0.849$. Similar to the pure Hubbard
or Holstein model the electronic kinetic energy of the AFM and CO state is larger
than the N (metallic) state, whilst the energy gain comes from the interaction
energies, $E_U$ for the AFM state and $E_g$ for the CO state, see
Fig. \ref{engUg_lamdepU25}, both for $U=2$ and $U=5$. The situation is
different for the BP state ($\lambda>3.1$ for $U=2$ and $\lambda>5.4$ for
$U=5$). Then the CO state wins through lower kinetic energy, whilst the
interaction energy is lower in the BP state. 

The phonon energy does not change much with $\lambda$ for values $\lambda<U$
in the N and AFM state, but increases rapidly for $\lambda>U$ both in the N
and CO state. This is mainly due to the increase in potential energy
$\omega_0^2\expval{\hat x^2}{}/2$, as will be seen in detail in Section
\ref{propphon} when we discuss $\expval{\hat x^2}{}$.

\noindent

The energy of the AFM state depends relatively little on the
electron-phonon coupling, i.e. if the order is not destroyed, the phonons have a
minor effect on the static properties. $E_U$ has a small value for weak
electron-phonon coupling and tends to $U/2$ in the case of strong coupling for
the BP and CO state. For weaker repulsion, $U=2$, the energies vary
continuously with $\lambda$. However, for the stronger 
coupling case $U=5$ we can clearly see the discontinuities at the AFM-CO
transition, $U\simeq \lambda$. $E_{\rm ph}$ jumps from a low value in the AFM
state to a large value in the CO state. Due to the increase in double
occupancy also $E_{U}$ increases suddenly at the transitions. Both of these
energetically unfavorable contributions for the CO state are counterbalanced
by the abrupt decrease of $E_{g}$, as the electron-phonon gives strong binding
energy in the CO state. As a result the total energy is continuous at the
transition, but shows kink.

For large $\lambda$ the interaction energy $E_g$ is proportional to
$-\lambda$. For the CO state this 
can be understood through a factorization of the expectation value in
$E_g$ in the electronic and phonon part. With equation (\ref{idxphico})
discussed later one finds then $E_g\to -4\lambda\Phi_{\rm co}^2$. Since in the
ordered state $\Phi_{\rm co}=1/2$ the behavior follows. 

\section{Quasiparticle properties in the normal state}

In a large part of the phase diagram in Fig. \ref{hfphasediag} near the AFM-CO
transition, the system without symmetry breaking is in the N metallic state,
which is a Fermi liquid. We can gain insight into the properties of the
system, when we analyze   the quasiparticle properties of the N state. The
states with broken symmetry can then be viewed as instabilities of the Fermi
liquid state. The renormalization factor $z$, which is related to the weight
of the quasiparticle peak and for a $\vk$-independent self-energy to the inverse of the
effective mass of the quasiparticles, $m^*/m_0=z^{-1}$. It can 
be calculated from the derivative of the self-energy as well as from the
analysis of the NRG low energy excitations at the fixed point. From the latter
procedure, one can also deduce a local effective quasiparticle interaction
$U^r$ by comparing the energy of the lowest two-particle excitation energy
$E_{pp}$ with the energy of two one-particle excitations $E_p$, $U^r\sim
E_{pp}-2E_p$. For details we refer to Ref. \onlinecite{HOM04,BH07c}. 

\subsection{Renormalized parameters as function of  $\lambda$}
\begin{figure}[!htbp]
\centering
\includegraphics[width=0.45\textwidth]{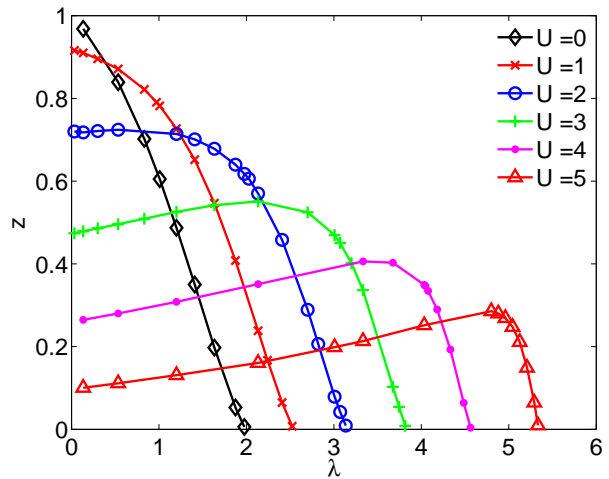}
\caption{(Color online) The $z$-factor for various $U$ as a function of $\lambda$.}       
\label{z_lambdadepdfU}
\end{figure}

First we discuss how the quasiparticle weight $z$ varies with $\lambda$, when $U$
is held fixed. 
It is shown for various values of $U$ as a function of $\lambda$ in
Fig. \ref{z_lambdadepdfU}. 

\noindent
We can identify different effects of the electron-phonon coupling. For $U=0$
and small values of $U$, increasing $\lambda$ leads to polaron formation and
localization of the charge carriers, which results in a reduction of $z$ and a
larger effective mass. Eventually, the metal-bipolaron transition is reached,
where $z\to 0$.  
For larger values of $U$ the electrons are already 
renormalized for $\lambda=0$ due to the Coulomb interaction. The first effect of increasing
the electron-phonon coupling is to reduce this effect and $z$ increases with
$\lambda$. Note that the effect is substantially less than what would be
expected for a pure Hubbard model with $U_{\rm eff}$. The maximal value
obtained occurs approximately when
$\lambda\simeq U$, i.e. the renormalization effects cancel there to the
largest extent leading to a least enhanced effective mass. Near the M-BP
transition $m^*/m_0$ diverges as for the weak $U$ case. Apart from the
approximate maximum no particular characteristic behavior is seen in $z$ near
the AFM-CO transition, $U\simeq \lambda$. 

\begin{figure}[!htbp]
\centering
\includegraphics[width=0.45\textwidth]{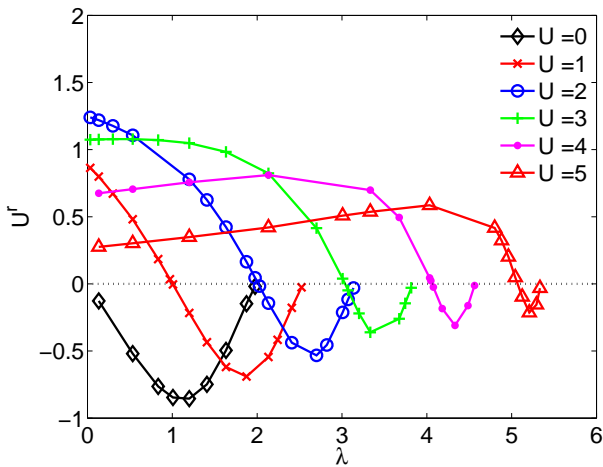}
\caption{The effective quasiparticle interaction $U^r$ for
  various $U$ as a function of $\lambda$.}        
\label{tU_lambdadepdfU}
\end{figure}

As shown in Fig. \ref{tU_lambdadepdfU} the effective quasiparticle interaction
$U^r$ varies between positive and negative values depending on $U$ and
$\lambda$. In the cases for finite $U$ it starts  
repulsive and goes to zero approximately where $U_{\rm eff}$ does. Then it
becomes negative such that there is an effective attraction between
quasiparticle excitations. There is a slight shift towards $\lambda>U$ for
this sign change to occur for larger values of $U$. This is in line with the
earlier observation that the CO state becomes the ground state when
$\lambda>U$.

The change of sign of  $U^r$ when $U\sim \lambda$ can be related to the
maximum found for $z$ at this point in
Fig \ref{z_lambdadepdfU}. At $U\sim \lambda$, $U^r\sim 0$, the
quasiparticles are effectively non-interacting. As $\lambda$ is varied from
this point, both $|U_{\rm eff}|$ and $|U_r|$ increase which causes a further
renormalization of the quasiparticles and a reduction in the value of $z$.
The decrease in $z$ from $U^r=0$ occurs irrespective of whether $U^r<0$, as in the pure
Holstein model, or $U^r>0$, as in the Hubbard model.

\noindent
Taking the viewpoint of instabilities of the  Fermi liquid,
one can infer that the ground state, AFM or CO, is 
determined by whether the low energy quasiparticles interact attractively or
repulsively.  The sign change in turn occurs when the  bare parameters
$U$, and $\lambda$ on high energies are equal. One might have expected that
in reducing the energy scale $\omega$ down to the phonon frequency $\omega_0$ that
the main  retardation effects would renormalize $U$ to some effective
value $\bar U$, where $\bar U$ would be of the order of $U^r$ in the pure
Hubbard model, which is such that $U^r\ll U$. The attractive term induced by the phonons would then 
contribute for $\omega \ll\omega_0$. The change of sign of the 
quasiparticle interaction would then be expected to occur when $\bar U\sim
\lambda$, which would correspond to a much smaller value of $\lambda$ than
$\lambda\sim U$. 
The fact that the transition and the change of sign of $U^r$ are
 found to occur when $U\sim \lambda$ indicates that $U$ term and the 
$\lambda$ term are renormalized in a similar way as the energy scale is
reduced.
 As noted before, however, when discussing the order parameter, the system
 cannot be described simply by an effective Hubbard model and both the
 electronic interaction $U$ and $\lambda$ play a role in determining the 
properties of the system in a certain phase. For instance, the CO
order parameter in Fig. \ref{phi_lamdepU5} 
corresponding to $U_{\rm eff}\simeq -0.01$ is $\Phi_{\rm co}\simeq 0.4$,
but the result would be close to zero for the pure Hubbard model with this
interaction on all energy scales.

\begin{figure}[!htbp]
\centering
\includegraphics[width=0.45\textwidth]{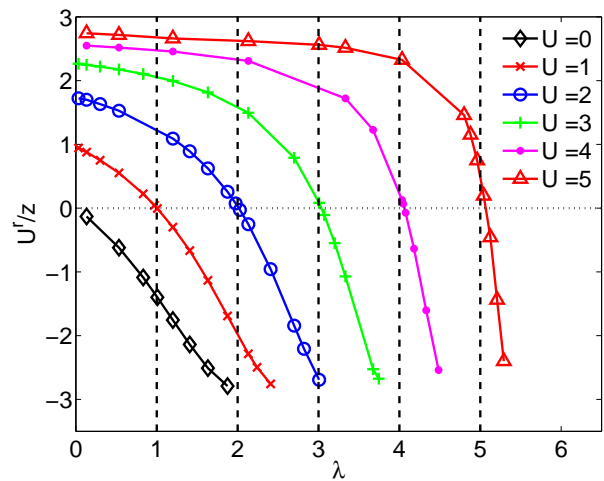}
\caption{The quantity $U^r/z$ for various $U$ as a function of $\lambda$.}       
\label{tUoz_lambdadepdfU}
\end{figure}

When $\lambda$ exceeds a certain value $|U^r|$ decreases again,
and we can also see that not only $z$ but also the effective quasiparticle
interaction $U^r$ goes to zero at the M-BP transition. It is of interest to
study the combined quantity $U^r/z=U^r m^*$, which is plotted in
Fig. \ref{tUoz_lambdadepdfU}. 

\noindent
This product of the effective quasiparticle interaction and effective mass takes
into account the aspect of the localization tendency of the quasiparticles as well as
their residual interaction. We can see that it shows a more universal
behavior for the different cases. It decreases monotonically as a function of
$\lambda$ for the given values of $U$. Close to the M-BP transition in all cases
$U^r m^*$ tends to a value between 2.5 and 3. One is therefore tempted
to identify this as the relevant quantity for an instability of the metallic
state, such that the M-BP transition occurs.

\subsection{Renormalized parameters as function of  $U$}
\begin{figure}[!htbp]
\centering
\includegraphics[width=0.45\textwidth]{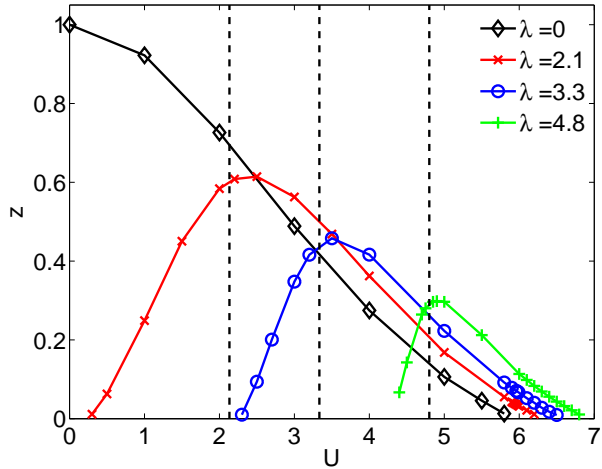}
\caption{The $z$-factor for various $\lambda$ as a function of $U$.}       
\label{z_Udepdfg}
\end{figure}

Similarly, we can also study the behavior of the quasiparticle properties for fixed
$\lambda$ as a function of $U$.
The quasiparticle weight $z$ is shown for various values of $\lambda$ as a function
of $U$ in Fig. \ref{z_Udepdfg}. 

\begin{figure}[!htbp]
\centering
\includegraphics[width=0.45\textwidth]{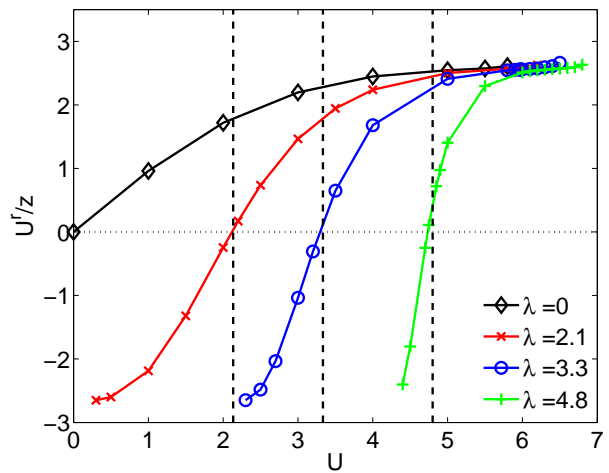}
\caption{The quantity $U^r/z$ for various $\lambda$ as a function of $U$.}       
\label{tUoz_Udepdfg}
\end{figure}

\noindent
For $\lambda=0$, $z$ is monotonically driven to zero when increasing $U$ to
the Mott transition. For finite values of $\lambda$ increasing $U$ leads to
an increase in $z$ when we come from the bipolaronic state and thus to a
de-renormalization as $|U_{\rm eff}|$ decreases. Similar to the case observed
above but more pronounced, 
one finds $z$ to be approximately maximal near the AFM-CO transition, $U\simeq
\lambda$, i.e. when $U_{\rm eff}$ as well as $U^r$ are close to
zero. Different renormalization effects have then canceled to maximal extent,
leaving relatively weakly renormalized, nearly non-interacting quasiparticles.
Again it is of interest to study the quantity $U^r m^*$, which is plotted in
Fig. \ref{tUoz_Udepdfg}. 

\noindent
As noted before it goes to zero where $U_{\rm  eff}$ does.   
Then it changes sign and increases with $U$ as long as the system is in the
metallic state. Close to the Mott transition it 
approaches a value 2.5-2.7 which is approximately the same for different
values of $\lambda$. It seems therefore that for this metal insulator
transition a universal value determines when it occurs.

\section{Properties of the phonons}
\label{propphon}
So far we have studied the CO-AFM transition via the electronic
properties, which where greatly influenced by the interaction with the bosonic
modes. In turn the properties of the local harmonic oscillator modes are also modified by
the coupling to the electronic degrees of the freedom, depending on the
different coupling strengths. First we consider  the phonon number expectation
value  $n_{\rm  ph}=\expval{b^{\dagger}b}{}$. 
It is expected to increase in the CO (BP) state as a high probability of local double occupation
leads to a charge redistribution and thus a (temporal) displacement of the lattice
ions. This means that phonons become excited, multiplied with $\omega_0$
this gives the energetic contribution $E_{\rm ph}=\omega_0 n_{\rm  ph}$ This
was discussed in Section \ref{difenergies} in Fig. \ref{engkinph_lamdepU25}
for $U=2$ and $U=5$ as a function of $\lambda$.

We find that the low value for $n_{\rm  ph}$ in the AFM state increases quite
slowly as the coupling strength is increased. Once the transition to the CO state
has occurred $n_{\rm  ph}$ increases substantially. The expectation values connect
continuously, but increase rapidly for larger $\lambda$. The value for the
normal state lies below the CO result but increases rapidly for $\lambda\simeq
3$, where the metal-bipolaron transition occurs. 
This behavior can be compared with the result for larger $U=5$, which is
shown in Fig. \ref{engkinph_lamdepU25} (bottom right). 
We can see that $n_{\rm  ph}$, being small in the AFM state, remains nearly
unaltered when the coupling is increased. The large $U$ and strong AFM order
suppresses local charge fluctuations. The value increases rapidly near 
$U_{\rm eff}=0$ in the CO state. The expectation value changes therefore
discontinuously at the transition. 
The behavior in the normal state is similar to the earlier case and the
metal-bipolaron transition ($\lambda\simeq 5.3$), which is also discontinuous
for these parameters.

\begin{figure}[!htbp]
\centering
\includegraphics[width=0.45\textwidth]{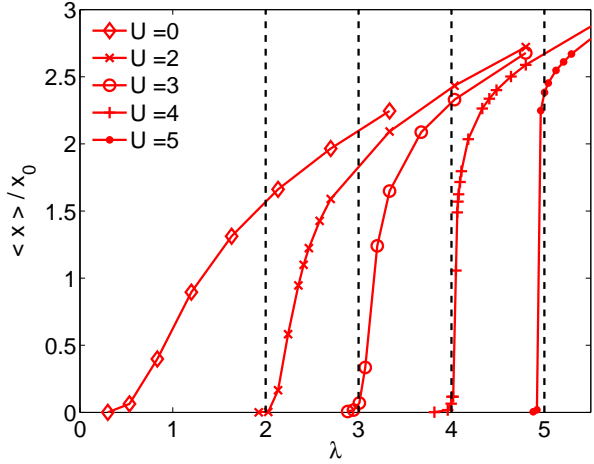}
\caption{(Color online) The expectation values $\expval{x}{}$ for various
  values of $U$ as a function of $\lambda$ for $\omega_0=0.6$ in the CO state.}       
\label{xexp_lamdepvarU}
\end{figure}

The charge order, which we have characterized by $\Phi_{\rm co}$, can also be
seen directly in the displacement expectation value on the $A$-sublattice
$\expval{\hat x_A}{}=\expval{b_A+b_A^{\dagger}}{}/\sqrt{2\omega_0}\equiv
\expval{x}{}$. This  value is always zero in the AFM and N state but finite
once the CO symmetry is broken. In Fig. \ref{xexp_lamdepvarU} it is plotted
for various values of $U$ as a function of $\lambda$.

\noindent
We can see that similar to the behavior of the order parameter $\expval{x}{}$
increases close to the transition, and more rapidly for larger $U$.
In its dependence on the coupling strength, it appears to be very similar to the order
parameter (Figs. \ref{phi_lamdepU2} and \ref{phi_lamdepU5}). In fact, one
can show that they are directly related by an exact identity, 
\begin{equation}
  \expval{x}{}=-\frac{2\sqrt{\lambda}}{\omega_0}\Phi_{\rm co},
\label{idxphico}
\end{equation}
which can be derived by considering an additional term
$H^i_c=\omega_0c(b_i+b_i^{\dagger})$ in the Hamiltonian and calculating the
derivative with respect to $c$. The numerical values for the left and right
hand side of equation (\ref{idxphico}) agree very well.
We can see that the slope at the transition increases with
$U$. At large $U$, or very small $\omega_0$, there is a very sharp transition as
seen for the case $U=5$.

\begin{figure}[!htbp]
\centering
\includegraphics[width=0.45\textwidth]{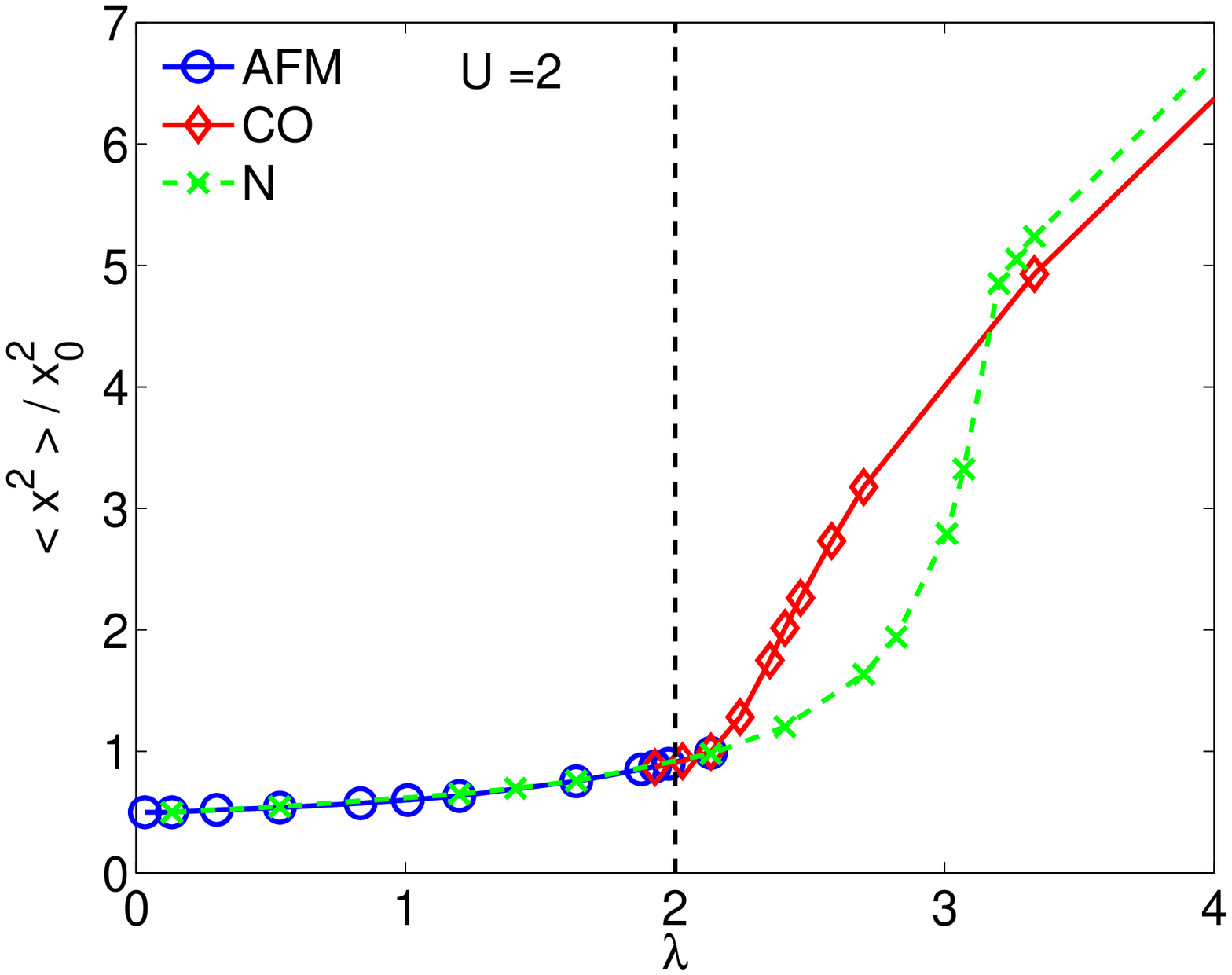}
\includegraphics[width=0.45\textwidth]{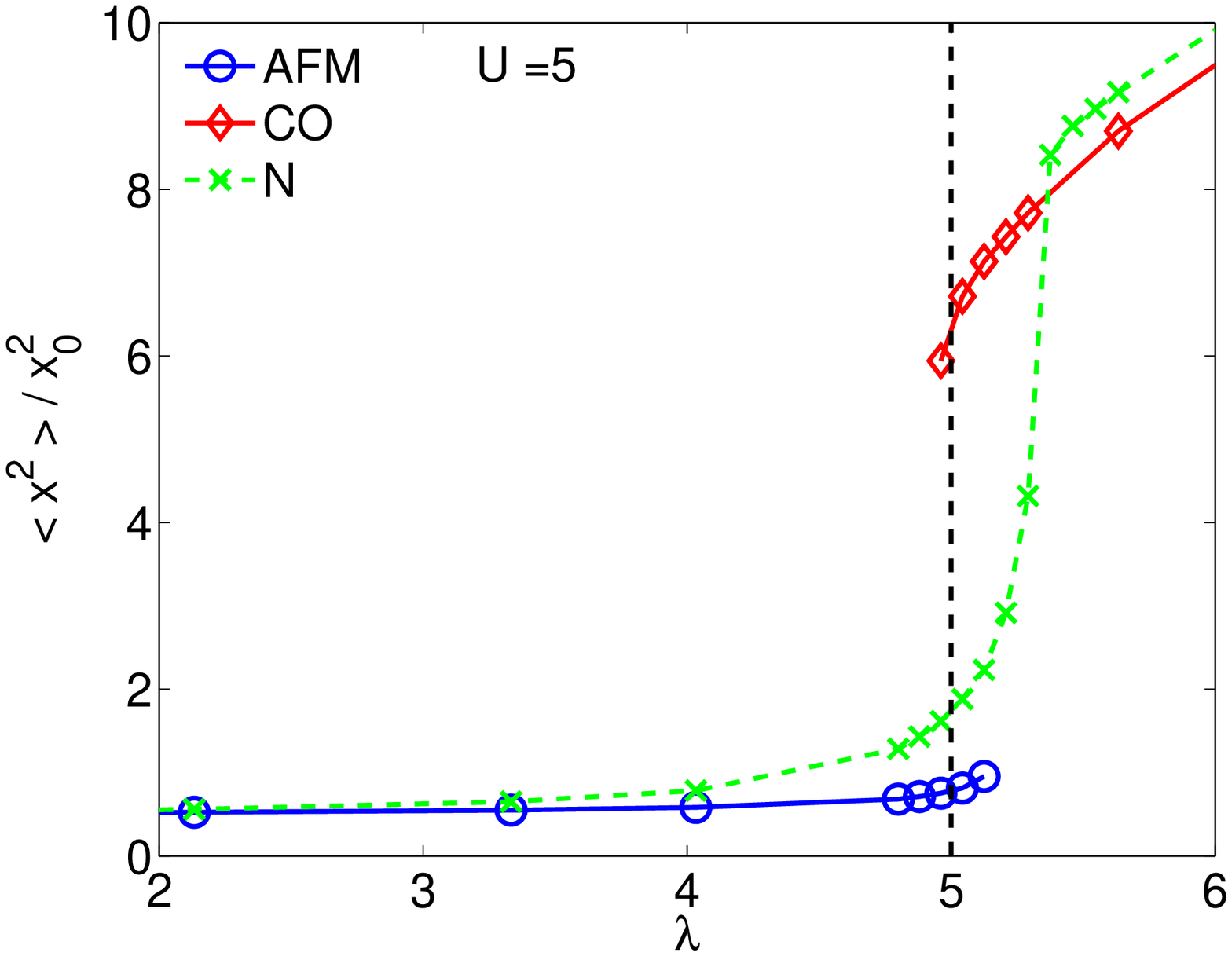}
\caption{(Color online) The displacement fluctuations $\expval{\hat x^2}{}$ for $U=2,5$ as a function
  of $\lambda$ for $\omega_0=0.6$ for the AFM, CO and normal state.}       
\label{x2ave_lamdepU25}
\end{figure}

The effect of the strong electron lattice coupling can be seen in the
displacement fluctuations $\expval{\hat x^2}{}$, which are plotted in
Fig. \ref{x2ave_lamdepU25} for $U=2$ and $U=5$ and a range of $\lambda$. 

\noindent
The behavior is reminiscent of the phonon expectation value $n_{\rm ph}$, where
a continuous rise near the transition is visible for weaker coupling and a
discontinuity from the AFM to the CO state of the stronger coupling case. Here
also the
normal BP state possesses a larger value than the CO state. 
The close comparison with the results for $n_{\rm ph}$ shows that for large
coupling in 
the BP and CO state $x_0^2 (1+4 n_{\rm ph})/2$ 
gives a good fit. This 
result can be derived by taking $n_{i,\uparrow}=n_{i,\downarrow}=1$, and
performing a displaced oscillator transformation to new phonon operators, $a$,
$a^{\dagger}$, $a^{(\dagger)}=b^{(\dagger)}+g/\omega_0$. The ground state
$|{\rm gs}\rangle$ then
corresponds to the state $a|{\rm gs}\rangle=0$, and in this state $\langle
b^{\dagger}b\rangle=g^2/\omega_0^2=n_{\rm ph}$, and  $\langle \hat
x^2\rangle=x_0^2(1+4g^2/\omega_0^2)/2$, giving the required result. The state 
$|{\rm gs}\rangle$ in the original basis corresponds to the coherent state
$\sum_n \frac{\alpha^n}{\sqrt{n!}} \ket{n}$  ($b^{\dagger}b\,\ket{n}=n\,\ket{n}$), with 
$\alpha=-\sqrt{n_{\rm ph}}$, and the result can alternatively be derived by
taking the expectation value of $\hat x^2$ in this state. The decoupled
oscillator state is  an eigenstate $\ket{n}$, but when strongly coupled to the
electronic system the nature of the state changes to the coherent ground state
 due to the displacement of the oscillator.

If we multiply $\expval{\hat x^2}{}$ by $\omega_0^2/2$ we obtain the potential
energy of the harmonic oscillator. The comparison with $E_{\rm ph}$ shows then
that most of the phonon energy is in the potential energy due to the charge
redistribution, and only small proportion in the kinetic energy of the oscillator.

\begin{figure}[!htbp]
\centering
\includegraphics[width=0.45\textwidth]{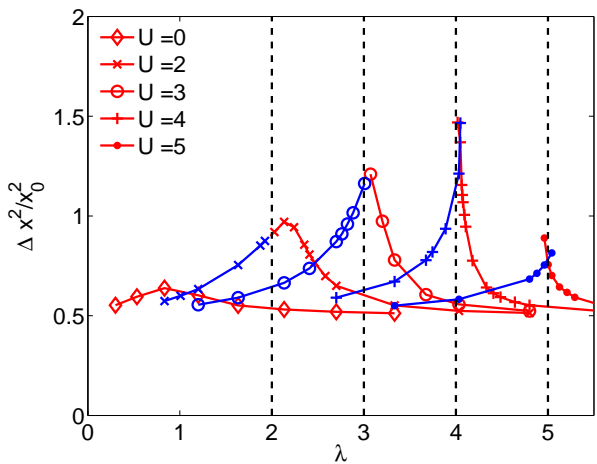}
\caption{(Color online) The expectation values $\Delta x^2$ for various
  values of $U$ as a function of $\lambda$ for $\omega_0=0.6$ in the AFM state
  for $\lambda<U$ and in the CO state for $\lambda>U$.}       
\label{x2mx2exp_lamdepvarU}
\end{figure}

The real lattice fluctuations are large in the BP state
where the local occupancy changes from double to zero, but $\expval{\hat
  x}{}=1$. A measure of these fluctuations is the quantity $\Delta 
x^2=\expval{(\hat x-\expval{\hat x}{})^2}{}=\expval{\hat x^2}{}-\expval{\hat x}{}^2$. This 
is a much smaller quantity than $\langle \hat x^2\rangle$ in the CO state and only large near the
transition, as can be seen in Fig. \ref{x2mx2exp_lamdepvarU}.  
In the uncoupled state  $\Delta x^2/x_0^2= \expval{\hat x^2}{}/x_0^2=1/2$.
In the displaced oscillator (coherent) state, which describes the strong 
coupling situation, 
$\langle \hat x\rangle^2=2x_0^2g^2/\omega_0^2$, so combining this with the
expression derived earlier for $\langle \hat x^2\rangle$,
 one again finds $\Delta x^2/x_0^2=1/2$.
Both limiting cases can be found in Fig. \ref{x2mx2exp_lamdepvarU}, where $\Delta
x^2$ increases with $\lambda$ in the AFM state. It then falls again
to $1/2$, when the system is strongly ordered. So the ``lattice fluctuations''
are largest at the transition.

It is possible to use the density matrix approach in the NRG \cite{Hof00}
together with the real space harmonic oscillator eigenfunctions to compute the
oscillator displacement probability function $P(x)$, which gives further insight into the
behavior of the phonons. Details for this method have been presented
elsewhere \cite{HB10}. The moments of this distribution function, $\expval{\hat
  x^m}{}=\integral{x}{}{}P(x)x^m$, can be calculated from $P(x)$ and are in
agreement with the value determined from the groundstate expectation
values.

\section{Spectral properties}
In this section we turn to the excitation spectra of the coupled
electron-phonon system. We first consider the electronic local lattice Green's function
$G_{\alpha,\sigma}(\omega)$, which is given by the momentum sum of the diagonal
element of (\ref{kgf}), and its spectral function
$\rho_{\alpha,\sigma}(\omega)=-\Imag G_{\alpha,\sigma}(\omega)/\pi$. In these
calculations the sublattice self-energies $\Sigma_{\alpha,\sigma}(\omega)$
enter, which are calculated according to Eq. (\ref{sefgf}) following
Bulla et al. \cite{BHP98}. We will comment on
complications which can arise in appendix \ref{selfencal}. $\rho_{\alpha,\sigma}(\omega)$ is of
special interest in studying  the broken symmetry behavior. In the normal state
the Green's functions are the same for the sublattices as well as the spin
projections. For the cases with symmetry breaking, CO and AFM, these function
will differ, i.e. for the CO case the sublattice Green's functions differ and
for the AFM case the different spin projections. We focus on the
$A$-sublattice majority spin spectral function
$\rho_{A,\uparrow}(\omega)$. Note that
at half filling the spectra for minority spin in the AFM case, $\rho_{A,\downarrow}(\omega)$,
and for the $B$-lattice for the charge order,
$\rho_{B,\uparrow}(\omega)$, can be obtained from $\omega\to -\omega$.
In order to calculate the full electron Green's function one has to put
the different sublattice Green's functions together \cite{BH07c}.

\subsection{Electron spectra}
\begin{figure}[!htbp]
\centering
\includegraphics[width=0.23\textwidth]{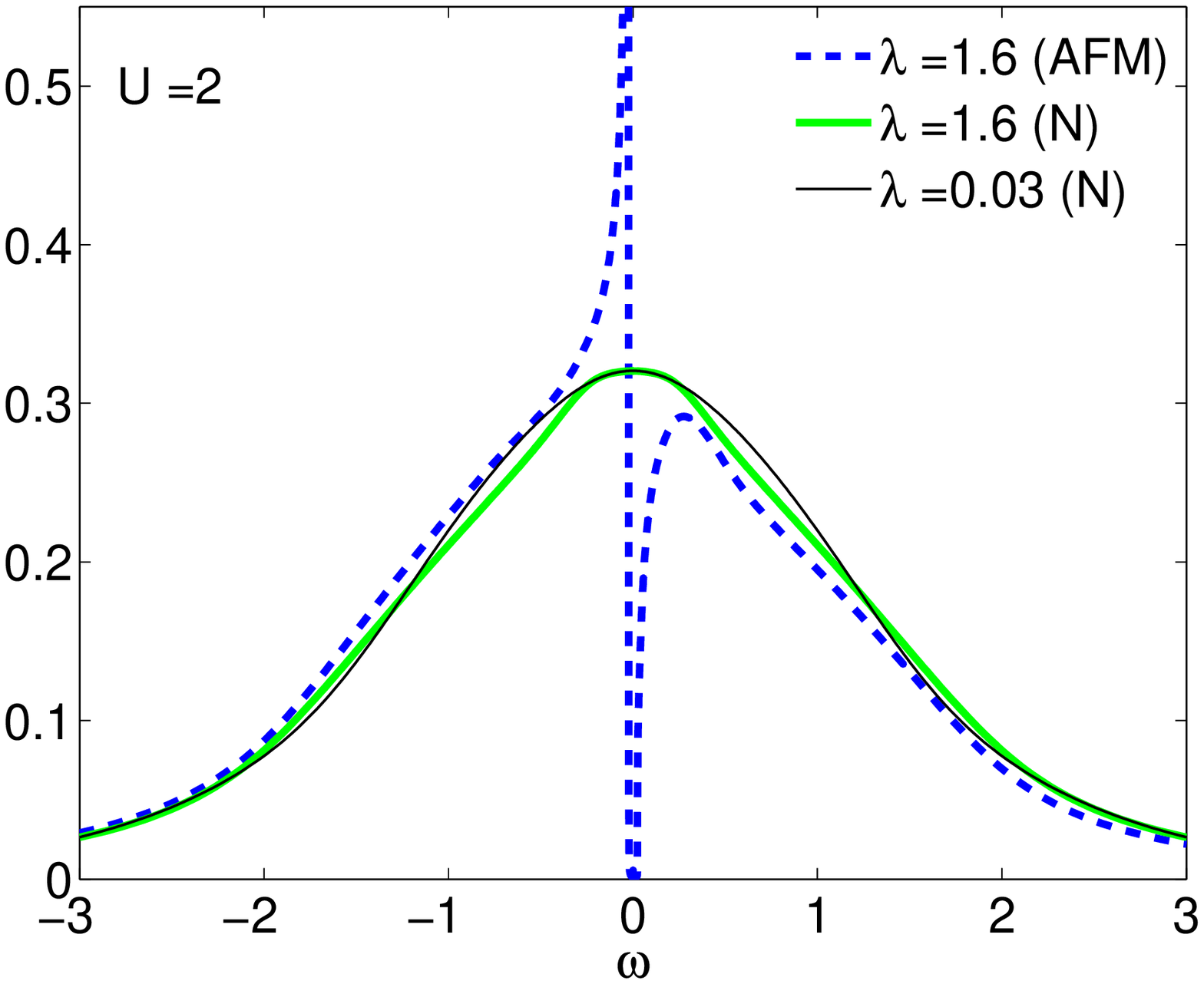}
\includegraphics[width=0.23\textwidth]{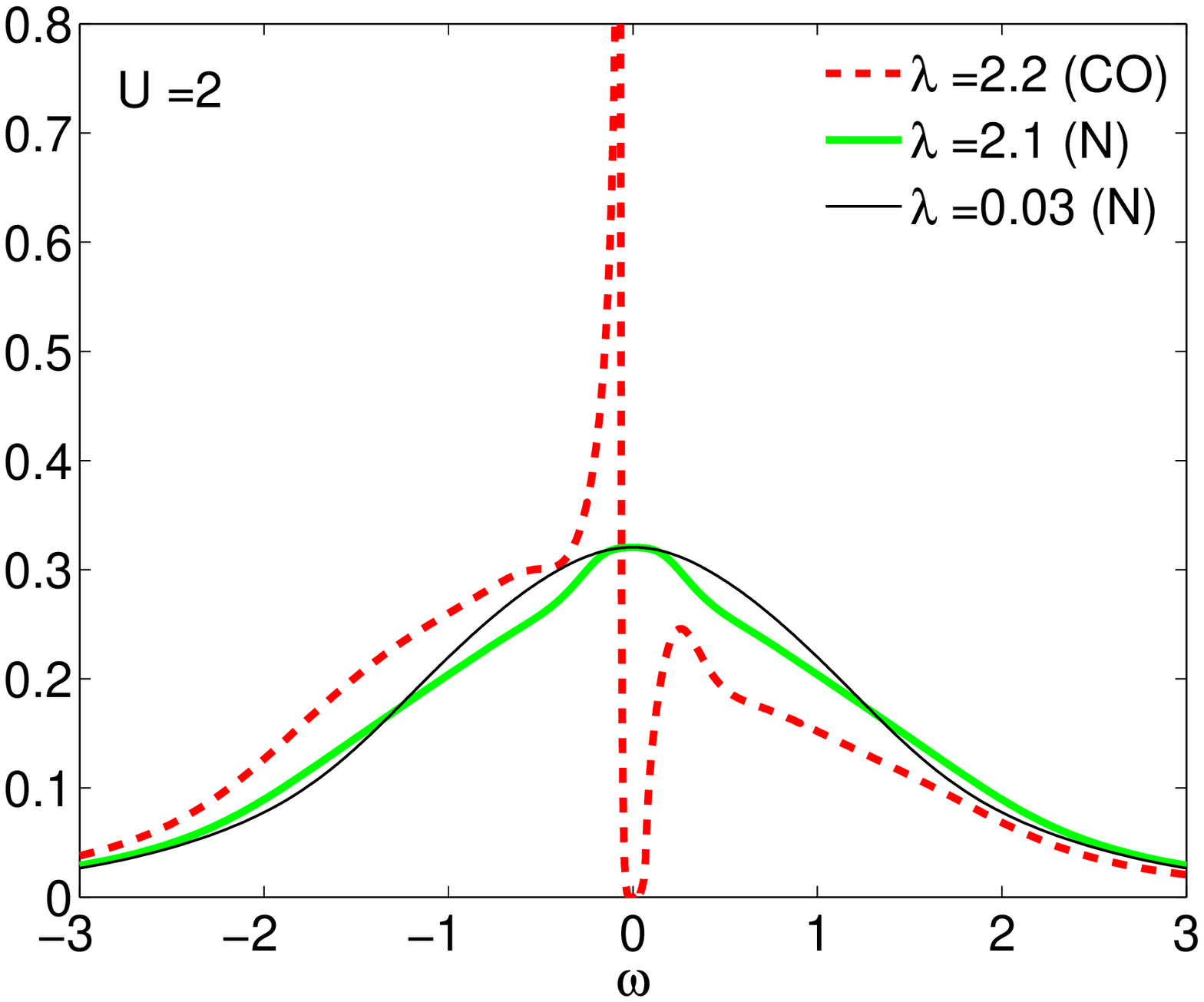}
\includegraphics[width=0.23\textwidth]{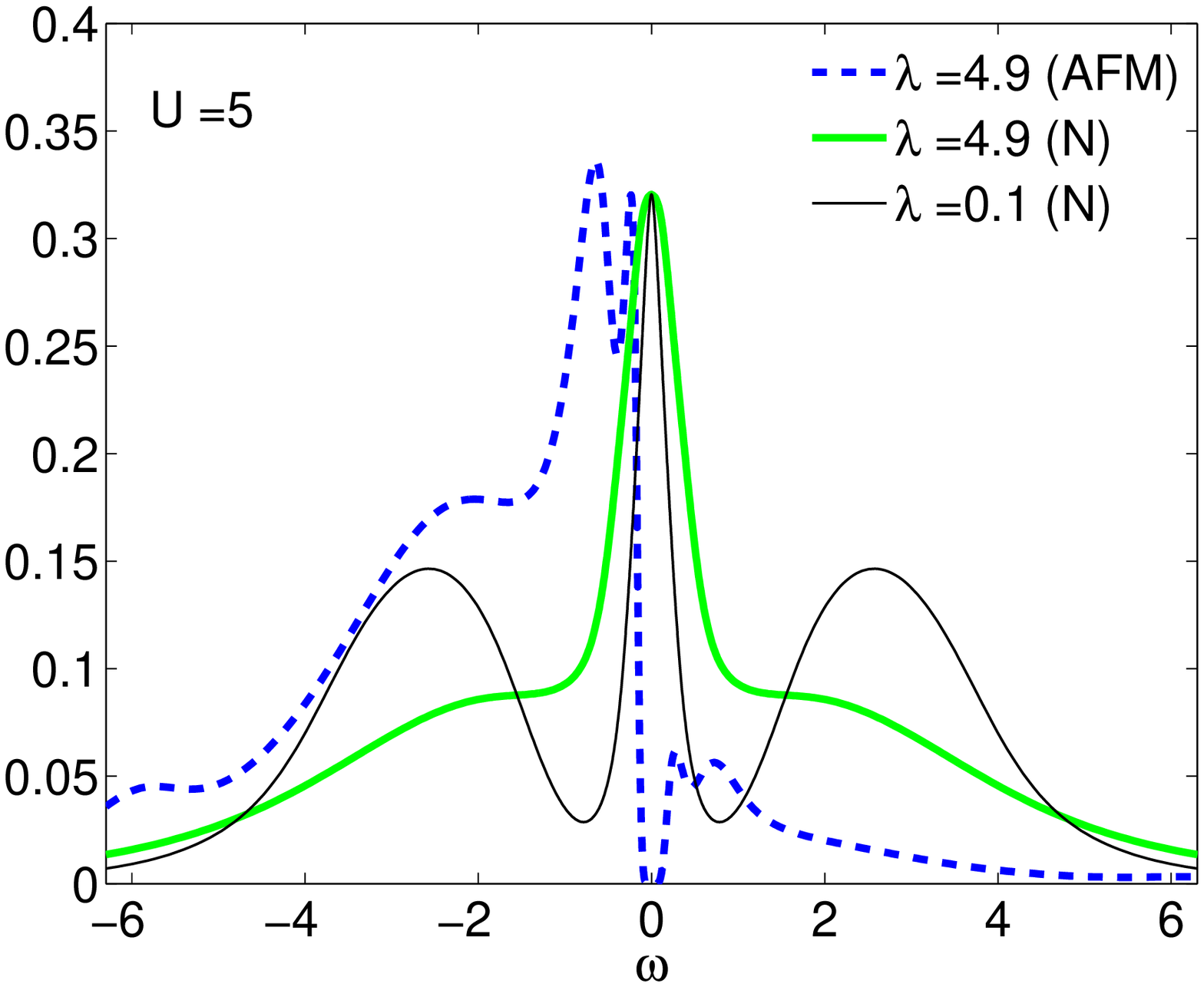}
\includegraphics[width=0.23\textwidth]{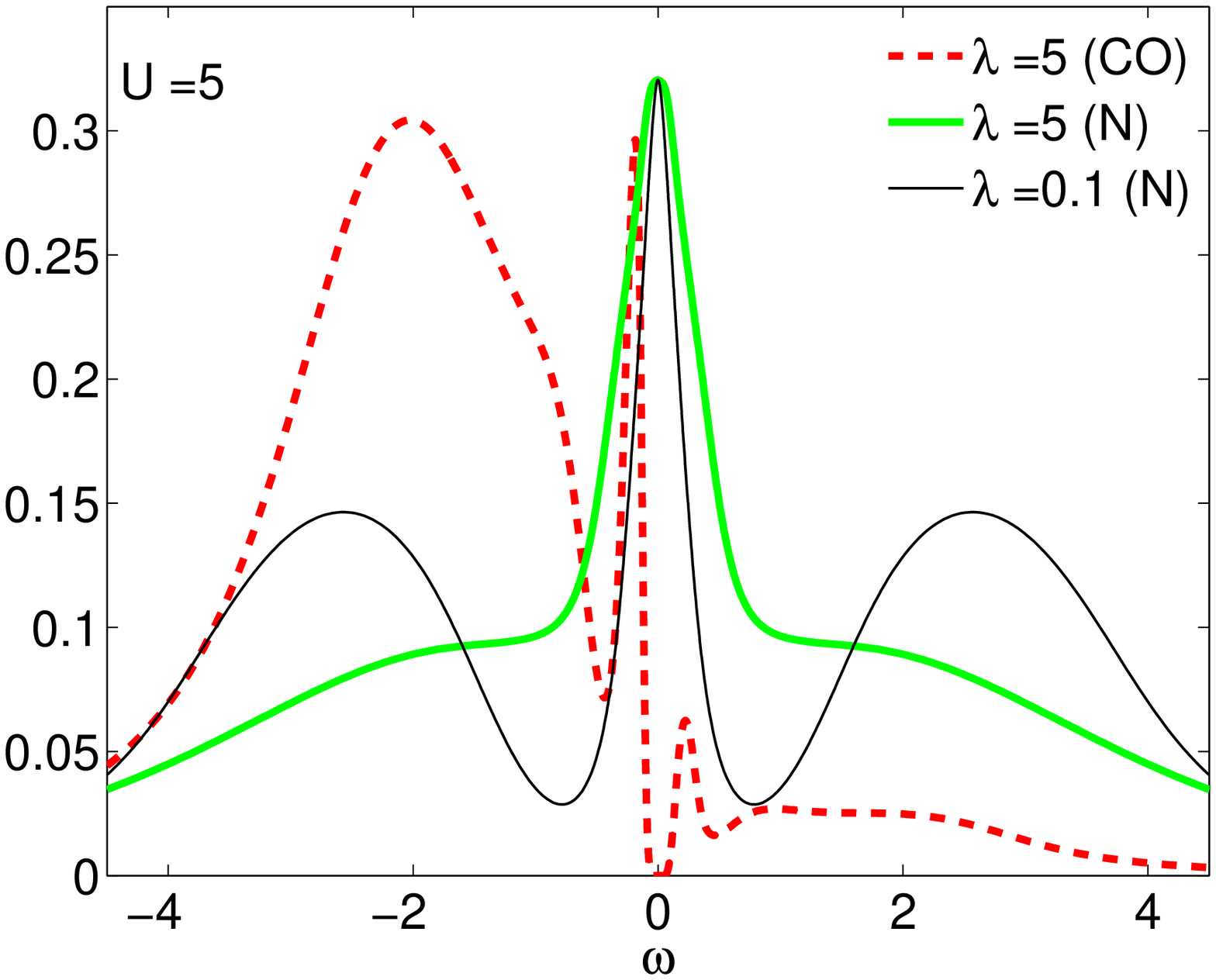}
\caption{(Color online) The local $A$-lattice spectral functions in comparison
  for $U=2$ (upper panel) and $U=5$ (lower panel). Left: Comparison of N state
  spectrum with AFM majority spin $\rho_{A,\uparrow}(\omega)$ near the
  transition. Right: Comparison of N state spectrum with CO $A$-site
  $\rho_{A,\uparrow}(\omega)$ near the  transition.}       
\label{elspec_neartransition}
\end{figure}

We first consider the electronic spectral functions near the transition for the
cases of $U=2$ and $U=5$. In Fig. \ref{elspec_neartransition} we plot the N
state spectral function in comparison with the corresponding symmetry broken
one. We have included a N state spectrum for $\lambda\to 0$ in order to first see
the effect of the phonons and associated modification in the normal state.

\noindent
For $U=2$ in the upper part of the figure we see that close to the transition
the N state spectrum deviates little from the $\lambda\to 0$ situation, with
only a small extra renormalization of the low energy excitations
[$z(\lambda=0)=0.73$ and $z(\lambda=2)=0.57$]; the metal bipolaron
transition with $z\to 0$ occurs for a larger $\lambda\simeq 3$. The imaginary
part of the electronic self-energy due to electron phonon scattering becomes
finite when $|\omega|>\omega_0^r$, where $\omega_0^r$ is the renormalized
phonon frequency (see Sec. \ref{phonspec}). 

For the AFM and CO symmetry broken state we can see characteristics of the
weak coupling instability at 
the Fermi surface ($\omega=0$) in the sublattice spectral function. 
It fits well to the mean field description, where a square root divergence is
found below the gap.\cite{ZPB02}
The higher energy parts are little modified for the case of $U=2$ apart from
the broadening of the band edges, but no features which can be attributed to
the phonons can be identified. 

In the lower panel of Fig. \ref{elspec_neartransition} we can see the
situation for $U=5$. For $\lambda\to 0$ we have the well-known three peak structure
with lower and upper Hubbard band and a quasiparticle peak in the N state. For
$\lambda\simeq 5$ the quasiparticle structure is still visible, but the
Hubbard peaks have been modified to high energy shoulders and cannot really
be recognized any more. The effect is as if the effective electron-electron
interaction is screened by the phonons. The quasiparticle
weight becomes larger in this regime on increasing $\lambda$ [$z(\lambda=0)=0.1$
and $z(\lambda=5)=0.27$, see also Fig. \ref{tU_lambdadepdfU}].
The low energy features of the spectra with symmetry breaking look
similar. Directly above and below the spectral gap  one sees 
pronounced peaks with larger weight for the one below the gap.  At
higher energies the spectra look different in both situations. 
Here both features from the large $U$ as well as from higher order polaronic
behavior can play a role. When interpreting the spectra one has to take into
account the broadening and the limited energy resolution of the NRG at higher
energies, which limit the accuracy. The AFM state with
strong electron phonon coupling seems to 
show polaronic behavior at multiples of $\omega_0$ with decreasing
weight. The charge ordered spectrum shows a principal peak at a position,
which is a bit less than the fully polarized mean field shift $U
n^{A}_{-\sigma}-2\lambda \Phi_{\rm co}\approx 2.5$. 
The large couplings for this case play a role on different energy scales.  
One may note that for the given parameters the order parameters
are rather large, $\Phi\approx 0.4$ (see Fig. \ref{phi_lamdepU5}), whereas the
spectral gap is relatively small. 

\begin{figure}[!htbp]
\centering
\includegraphics[width=0.23\textwidth]{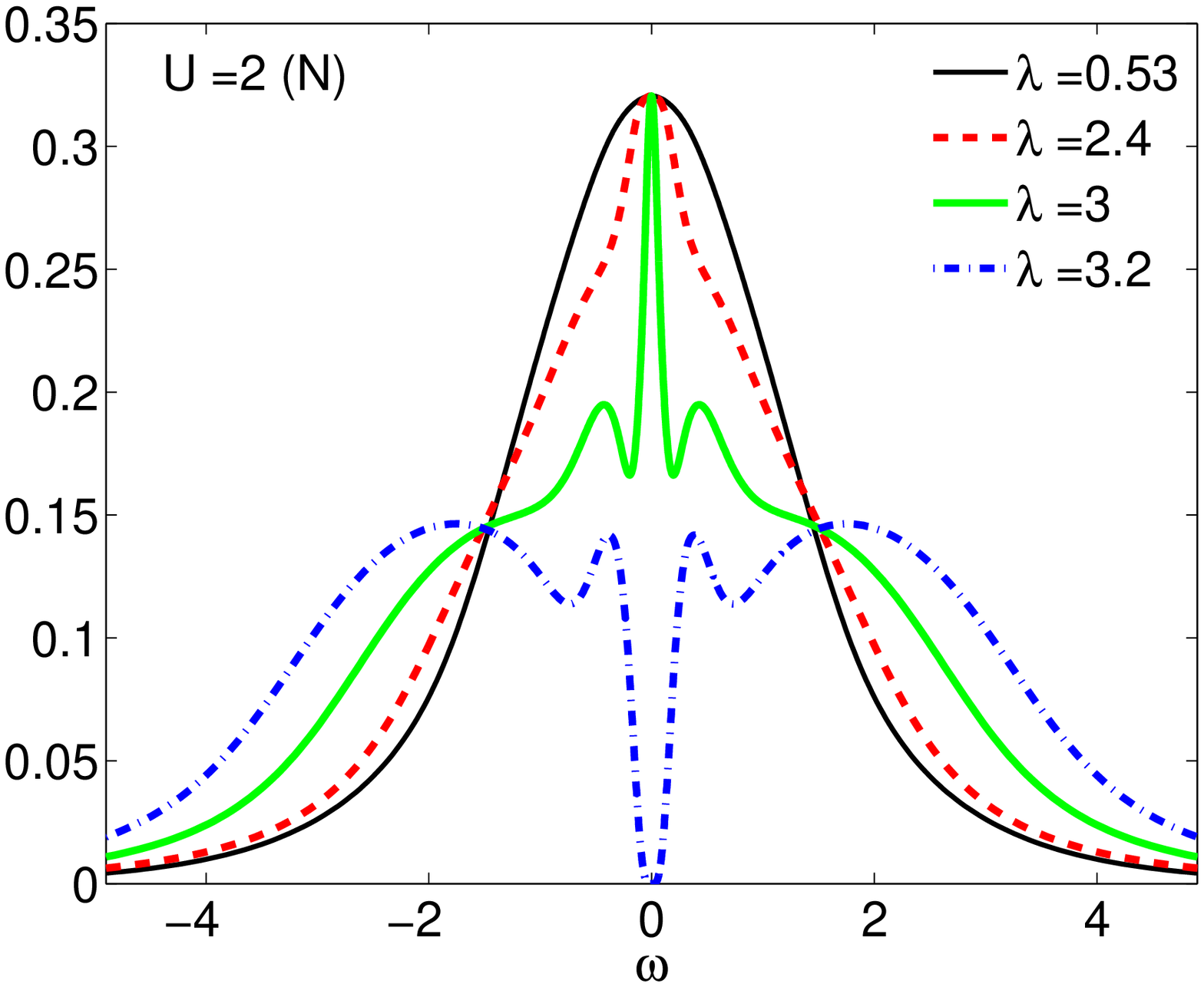}
\includegraphics[width=0.23\textwidth]{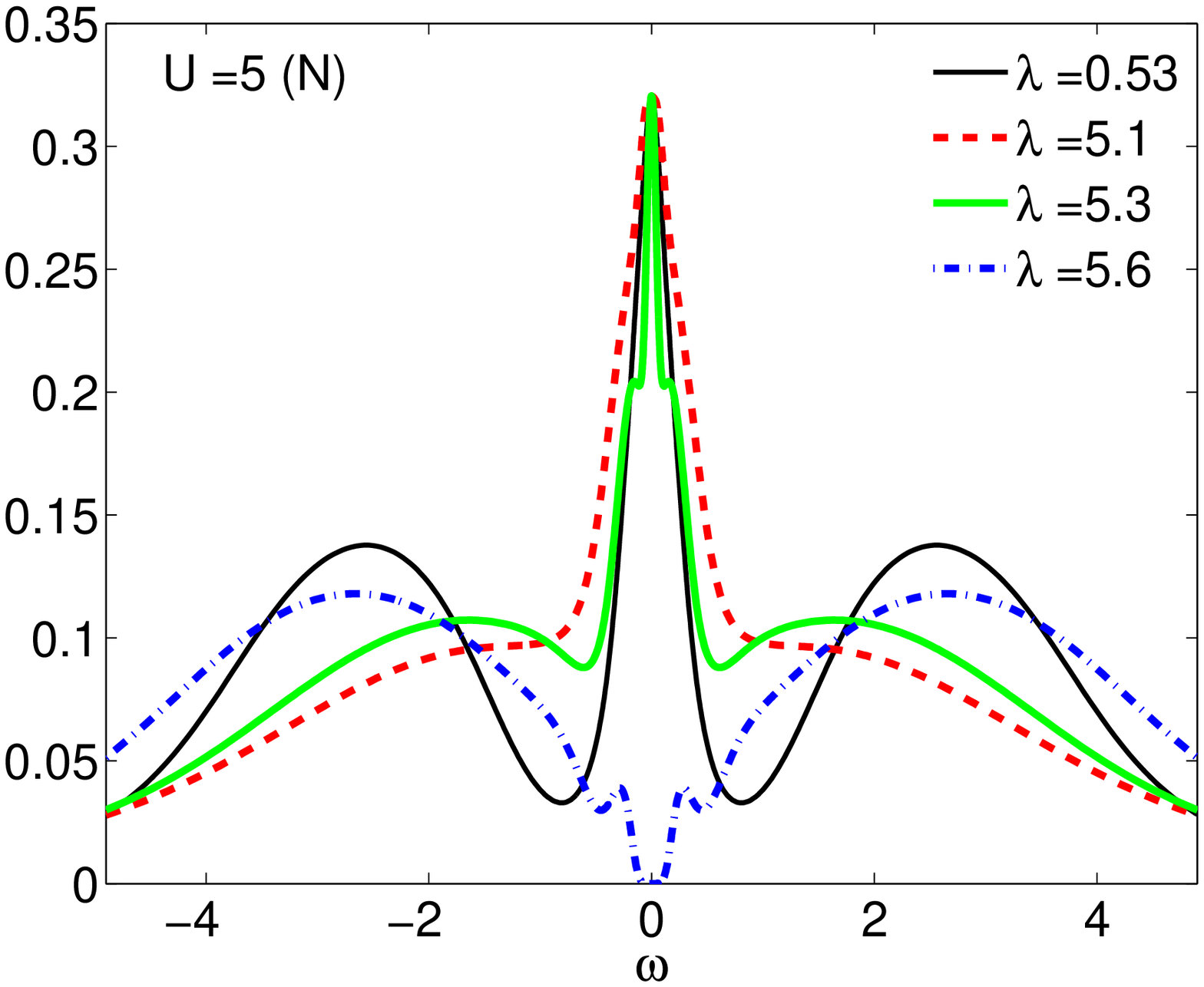}
\includegraphics[width=0.23\textwidth]{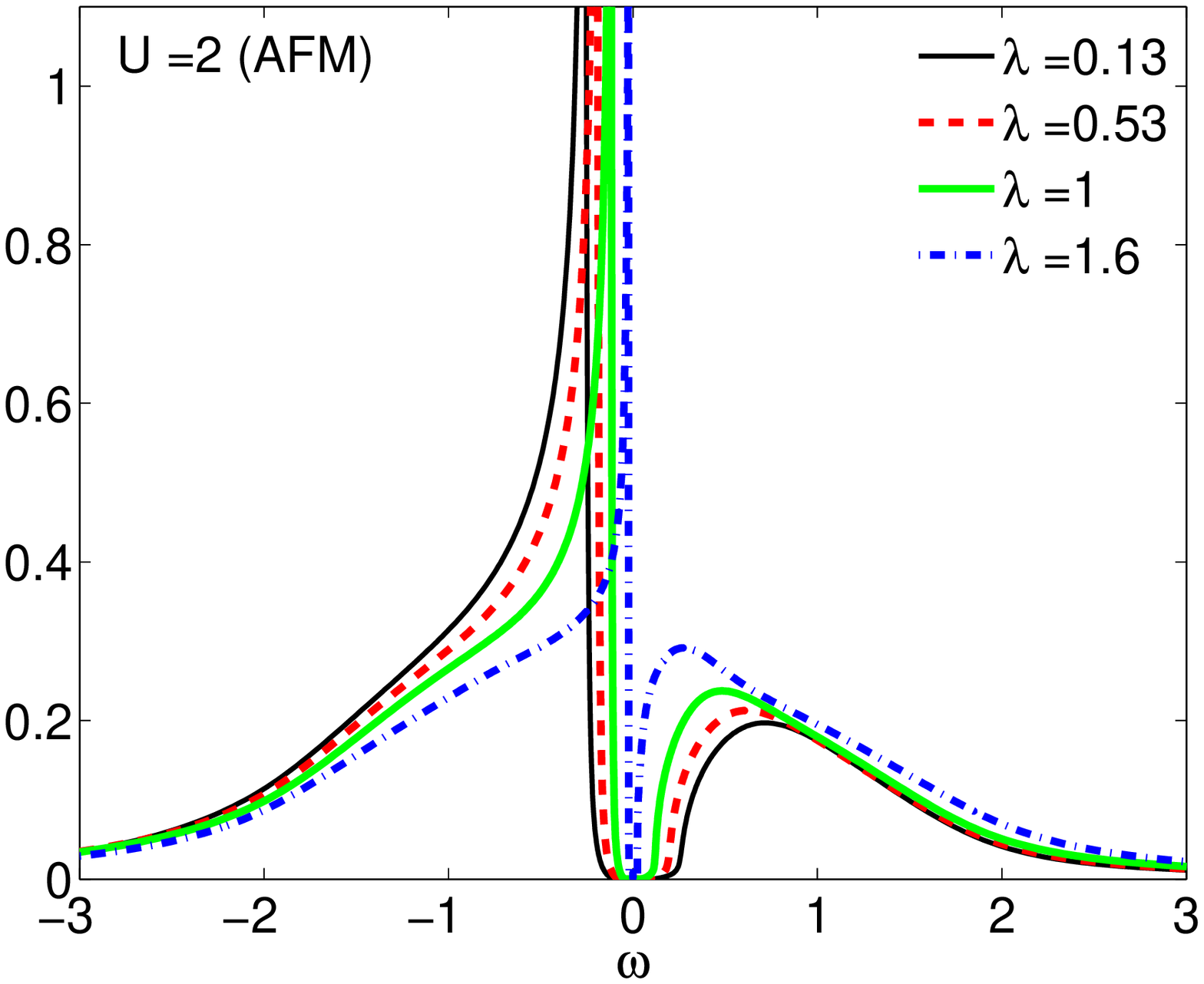}
\includegraphics[width=0.23\textwidth]{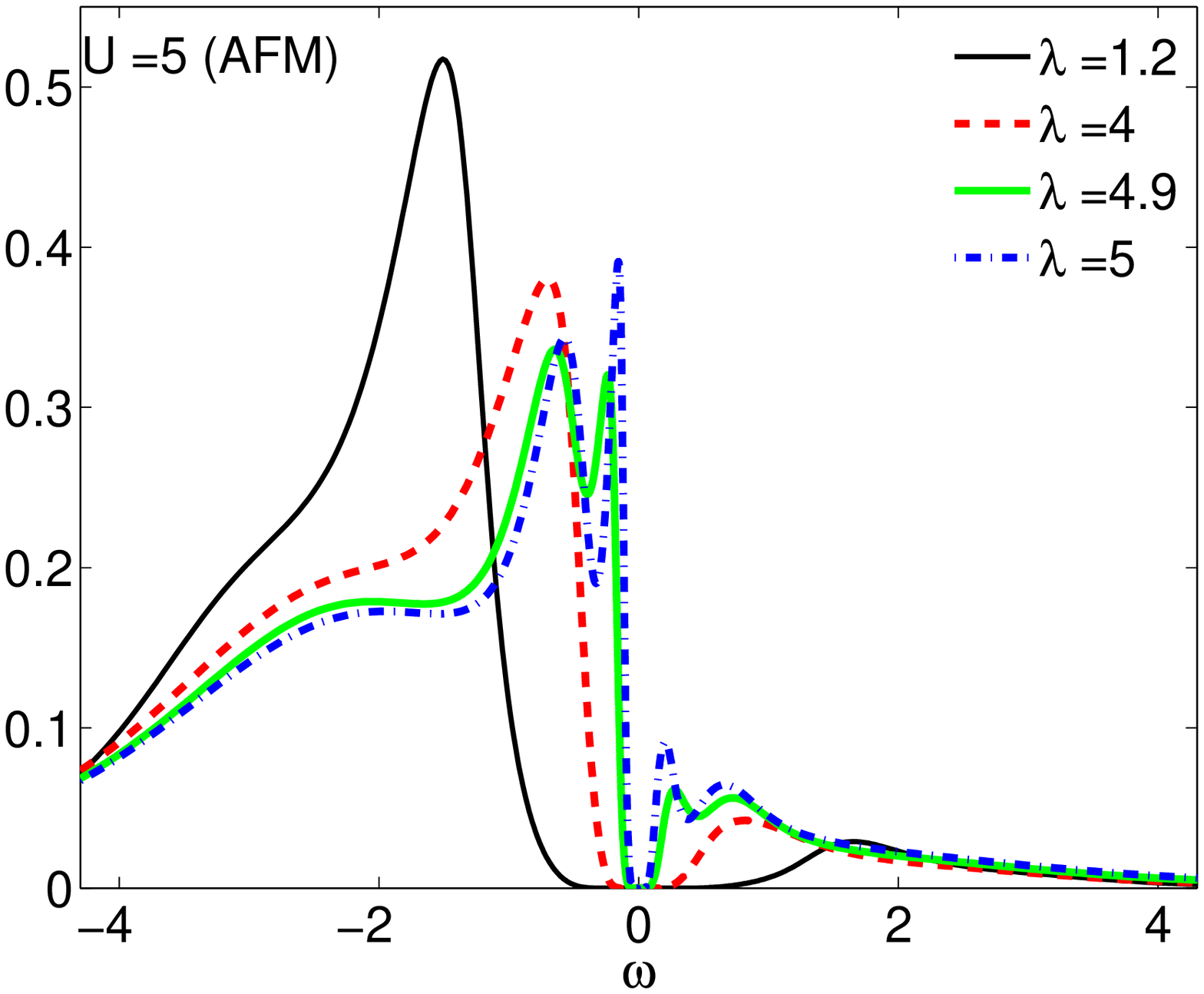}
\includegraphics[width=0.23\textwidth]{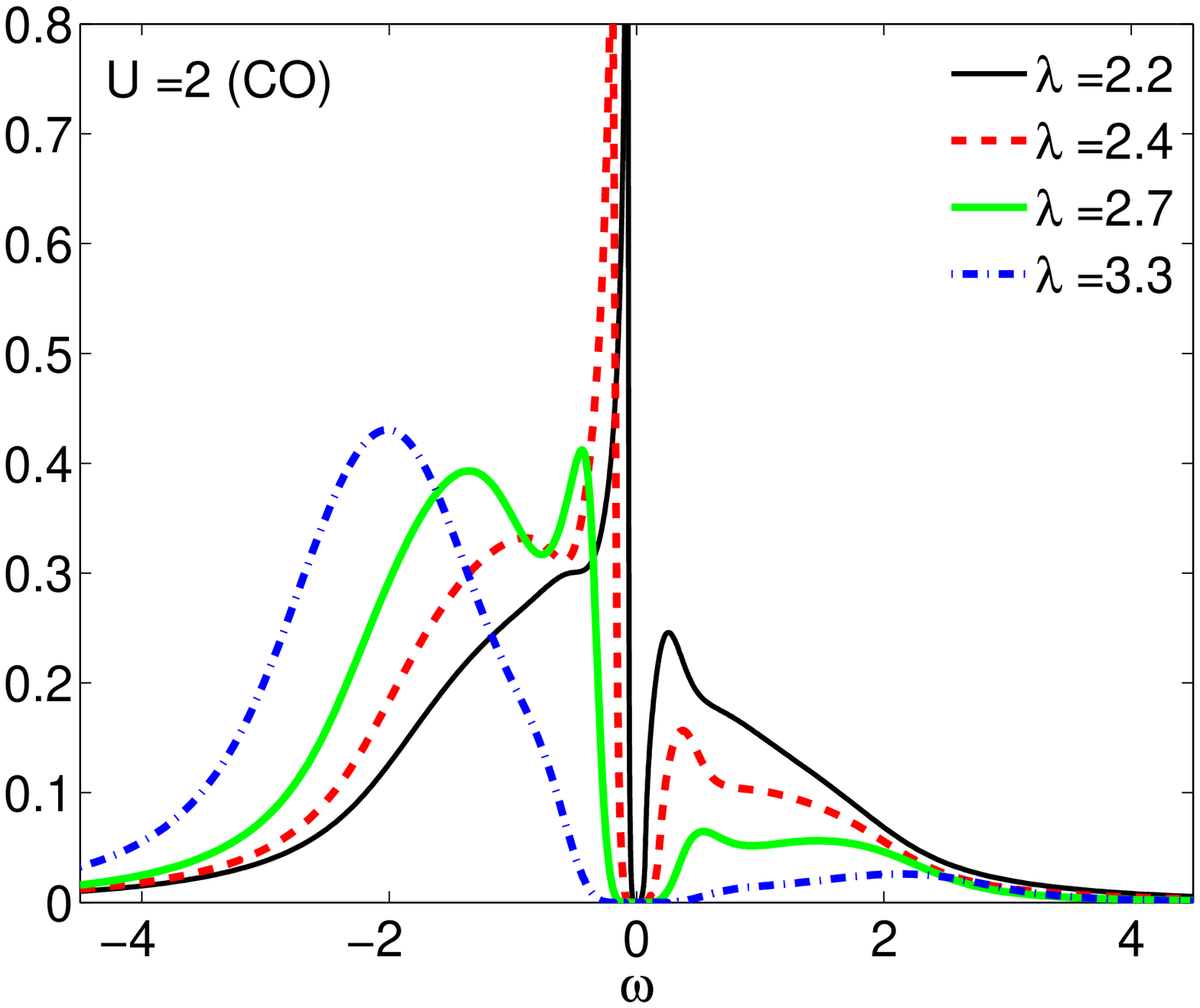}
\includegraphics[width=0.23\textwidth]{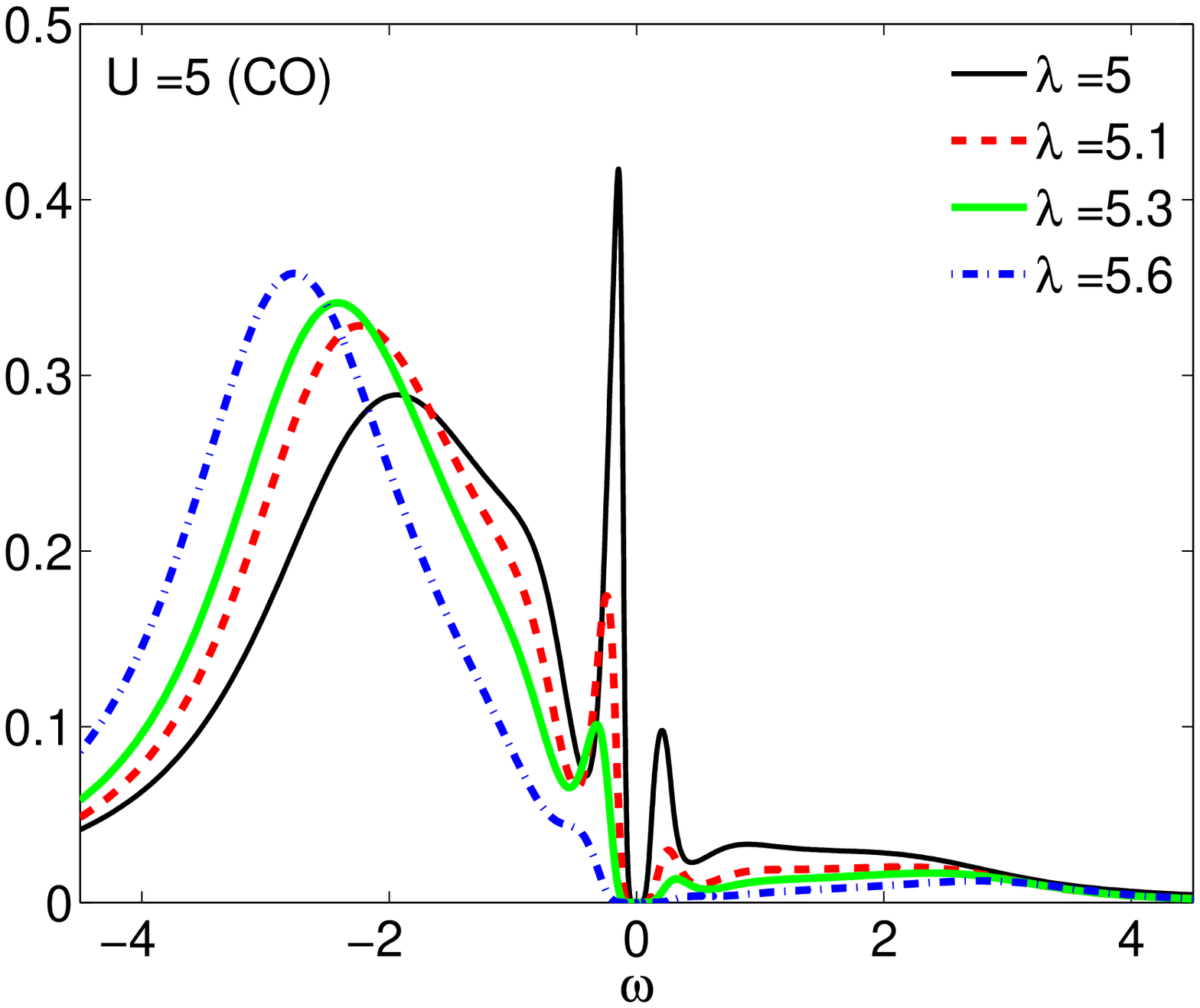}
\caption{(Color online) The local $A$-lattice spectral functions in comparison
  for $U=2$ (left) and $U=5$ (right) for various $\lambda$ and
  $\omega_0=0.6$. Top panel: N state spectrum. Middle panel: AFM
  majority spin $\rho_{A,\uparrow}(\omega)$. Lower panel: CO $A$-site
  $\rho_{A,\uparrow}(\omega)$.}       
\label{elspec_varlambdaU25}
\end{figure}

To study the behavior of the spectral functions at different couplings we give
various plots in Fig. \ref{elspec_varlambdaU25}. On the left hand side we keep
$U=2$ fixed and show (from top to bottom) the spectra for N, AFM and CO state
for various $\lambda$. On the right hand side $U=5$ is kept fixed.

\noindent
The top panel shows the metal to bipolaron transition when $\lambda$ is
increased. One can see the strong narrowing of the quasiparticle band near the
transition, which is accompanied by $z\to 0$. A spectral gap develops when
$\lambda$ exceeds a critical coupling. The details of the spectral functions
have been analyzed by Koller et al. \cite{KMH04}. 
In the AFM case in the middle we can see for the weaker coupling case how, on
increasing $\lambda$, the AFM order and magnitude of the spectral gap decreases. The electron
phonon coupling is effective here in screening the repulsive $U$-term. No polaronic
features can be identified in the spectra as the coupling is fairly weak. At
stronger coupling, the AFM state is hardly affected for a range of
$\lambda$. When approaching the transition we find visible modifications of
the spectral functions including a reduction of the spectral gap and polaronic
peaks.  

In the bottom part the spectra in the CO state can be seen. Near the
transition the spectra differ for the weak and strong coupling case as
discussed before, but when $\lambda$ exceeds $U$ by a certain amount the
spectra look very similar, and the different $U$ term is not directly visible
anymore. As noted before the main peak for the sublattice spectra is located near the 
mean field shift and its position moves linearly with $\lambda$ as
expected. There is a pronounced quasiparticle peak near the transition which
becomes suppressed for larger values of $\lambda$. This suppression can be
partly due to the broadening in the NRG procedure as discussed in detail for
superconducting solutions \cite {BHD09}.

\subsection{Phonon spectra}
\label{phonspec}
In this section we study the spectral properties of the phonons. From these we
can find out how the excitations of the bosonic sector are modified through the interaction with
the electronic system. Especially near a transition a strong phonon softening
can be indicative of a lattice modification or instability. 
We consider the function 
\begin{equation}
B(\omega)=\gfbraket{b;b^{\dagger}}_{\omega},
\end{equation}
which can be calculated in the NRG from the matrix elements and
excitations. The spectral function, $\rho_b(\omega)=-\Imag B(\omega)/\pi$, 
has the properties  at $T=0$,
\begin{equation}
\integral {\omega}{-\infty}{\infty}\rho_b(\omega)=1,\;\;\;
\integral {\omega}{-\infty}{0}\rho_b(\omega)=-n_{\rm ph},
\label{sumrules}\end{equation}
and the free propagator has the form
\begin{equation}
  B^0(\omega)=\frac1{\omega^+-\omega_0}.
\end{equation}
We define a phonon self-energy $\Sigma_{\rm
  ph}(\omega)$ the full propagator reads
\begin{equation}
B(\omega)=\frac1{\omega^+-\omega_0-\Sigma_{\rm  ph}(\omega)}.
\end{equation}
For a decoupled electron-phonon system $\Sigma_{\rm  ph}(\omega)=0$ such that
$\rho_b(\omega)$ is a delta function peaked at $\omega_0$. In the strongly
interacting system the mode can be renormalized to $\omega_0^r$ and
broadened, however, no $\vq$-dependence develops in the infinite dimensional
model. 

\begin{figure}[!thpb]
\centering
\includegraphics[width=0.23\textwidth]{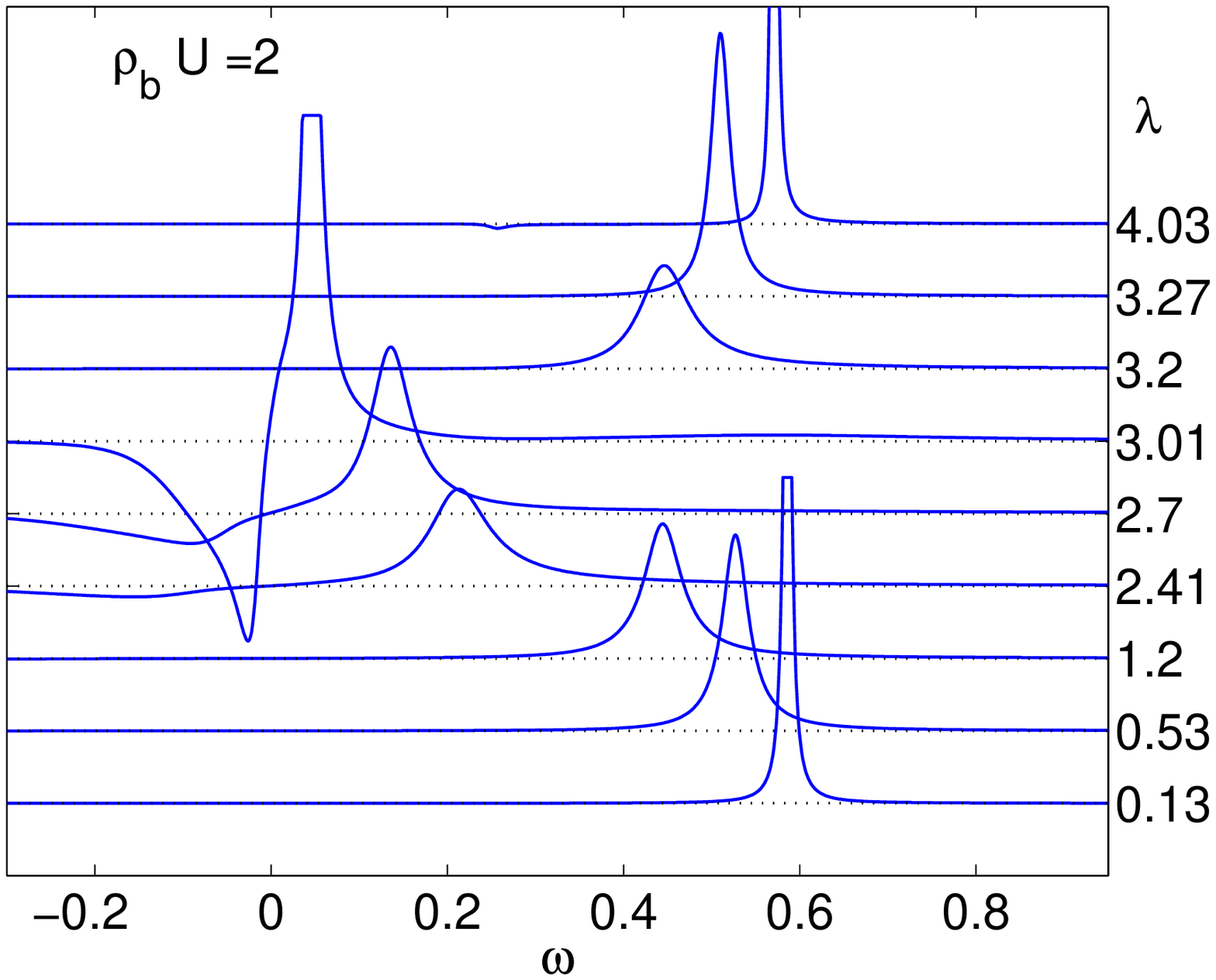}
\includegraphics[width=0.23\textwidth]{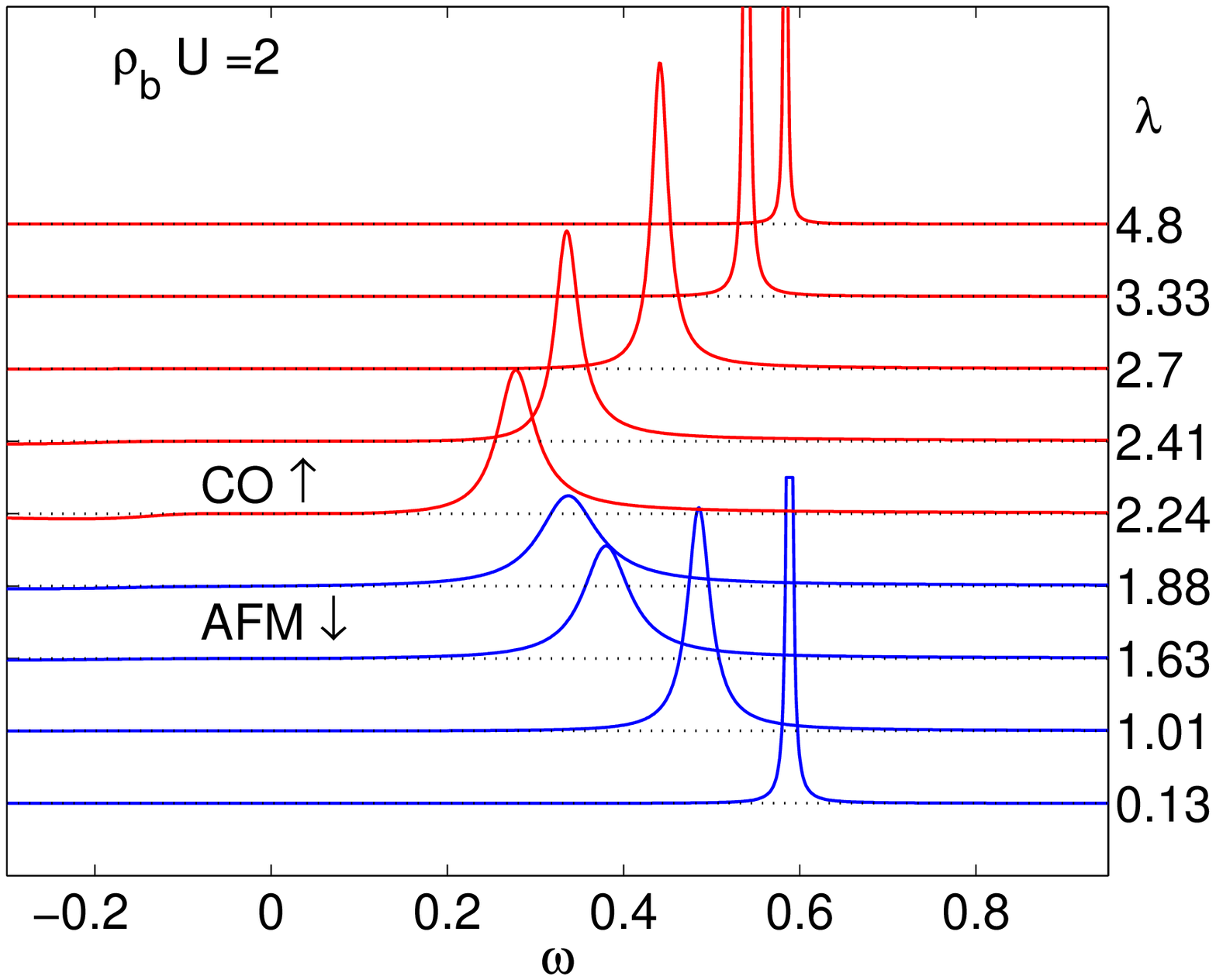}
\includegraphics[width=0.23\textwidth]{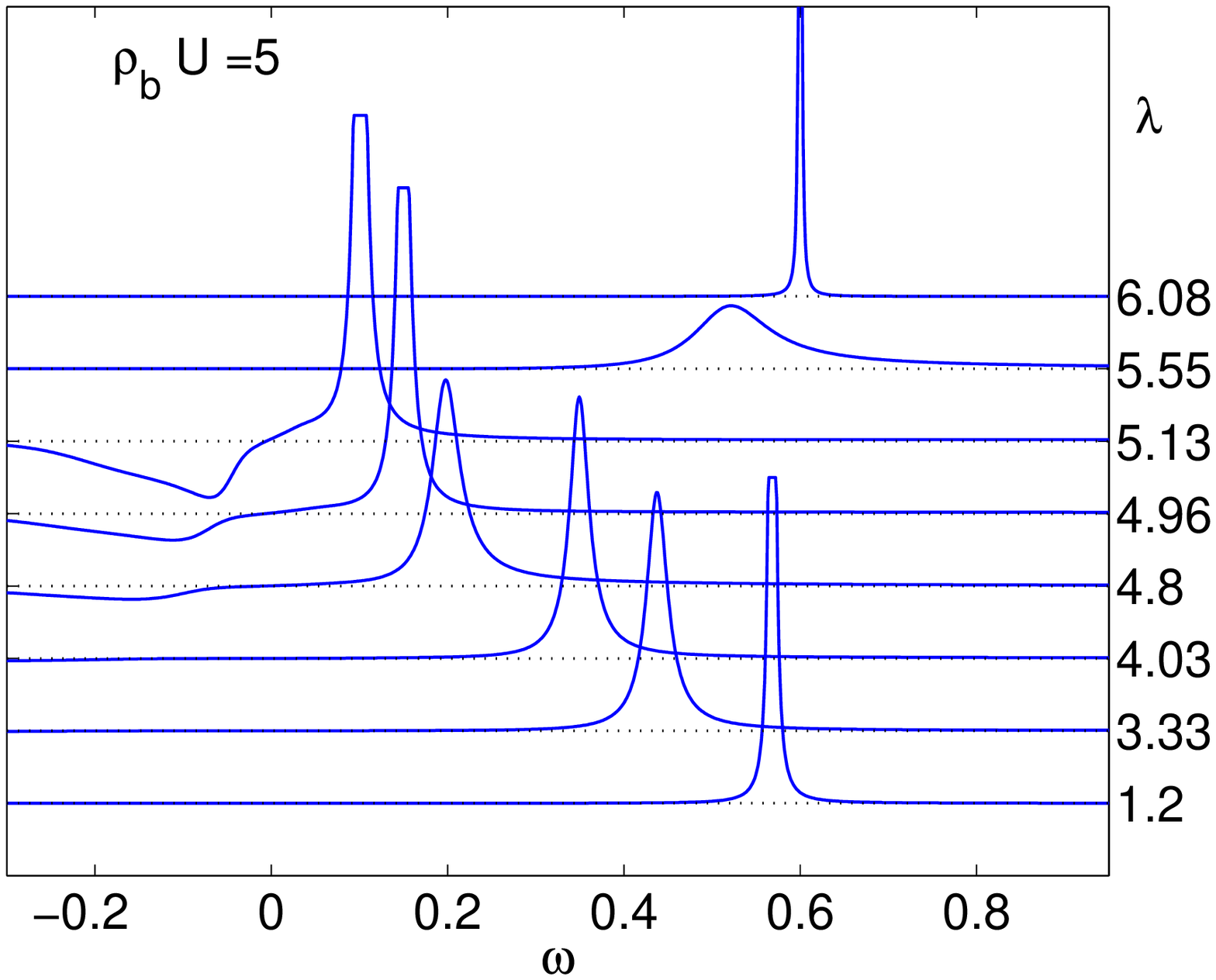}
\includegraphics[width=0.23\textwidth]{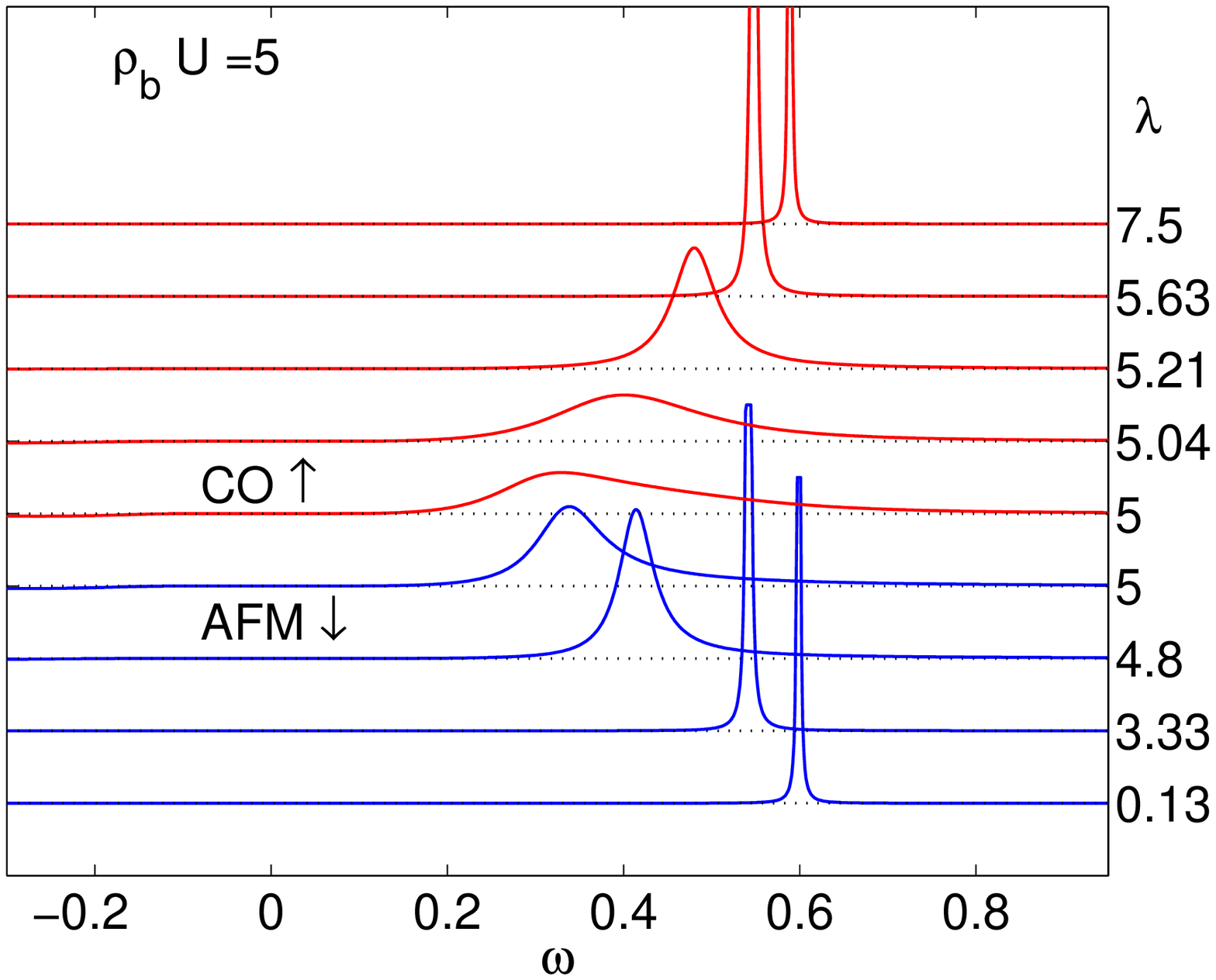}
\caption{(Color online) The local phonon spectral functions $\rho_b(\omega)$
  in comparison   for $U=2$ (upper panel) and $U=5$ (lower panel). Left:
  N state. Right: AFM state (smaller values of $\lambda$ and CO state for the
  larger values of $\lambda$.}       
\label{phonspec_neartransition}
\end{figure}

Again we focus on the cases with fixed $U=2$ and $U=5$ for variable $\lambda$, and
we compare the results from the N state with the ordered state. In
Fig. \ref{phonspec_neartransition} we plot in the upper part the results for
the spectral function of the phonons $\rho_b(\omega$) for
$U=2$ for the N state (left) and the ordered states (right) and the lower part
for $U=5$. 

When the electron and phonon systems are weakly coupled the expected
delta-function for $\rho_b(\omega)$ is found.  When $\lambda$ is increased the
phonon mode is substantially renormalized and softens, most markedly at the
M-BP transition for $\lambda\simeq 3$ where $\omega_0^r\to 0$. The negative
spectral weight for 
$\omega$, which builds up there, is directly related to the phonon expectation
value $n_{\rm ph}$. In the BP state it is not resolved any more. The phonon
mode then hardens back to $\omega_0$. Apart from the softening we also find a
broadening of the phonon spectrum, when the system is strongly coupled. The
behavior for $U=5$ is similar to the $U=2$. A more detailed discussion can be
found in Ref. \onlinecite{KMH04}.

\noindent
In the right hand side panels of Fig. \ref{phonspec_neartransition}   the
 corresponding behavior of $\rho_b(\omega)$
is shown  in the
states with broken symmetry. The results for $\lambda<U$ correspond
to the AFM phase and the ones for $\lambda>U$ to the CO phase. Comparing
the metal-bipolaron transition to the AFM-CO transition we find that the
effect of the phonon softening is much reduced. For $U=2$ we can see a visible
effect that the oscillator mode is renormalized to $\omega_0^r\simeq 0.27$ and
broadened. The effect is comparable to the normal state for the same values of
$U$ and $\lambda$, but  not as strong as at the M-BP transition. The
softening is of similar magnitude in the stronger coupling case $U=5$. There a
significant broadening of the phonon mode visible near the 
transition due to the large coupling of the phonons to the electronic
system. When the system is well in the ordered state, AFM or CO, the phonon
dynamics is little modified by the electronic system and a nearly free phonon
mode is observed both for large $U\gg \lambda$ in the AFM state and $U\ll
\lambda$ in the CO state. The sum rules for $\rho_b(\omega)$ given in equation
(\ref{sumrules}) are satisfied to within a few percent for smaller values of
$\lambda$. For larger $\lambda$ in the BP and CO state the sum rules are not
well satisfied due to reasons discussed in Ref. \onlinecite{KMH04}.

\section{Conclusions }
In this study of the competing interactions in the Holstein-Hubbard model
we have examined the transitions to both AFM and CO, both in the weak and
strong coupling regimes. To lowest order, the effective frequency dependent
interaction  $U_{\rm eff}(\omega)$ between the electrons is given by
\begin{equation}U_{\rm eff}(\omega)=U+{2g^2\omega_0\over \omega^2-\omega_0^2},
\end{equation}
the second retarded term arising from phonon exchange.
On the lowest energy scale $\omega=0$, $U_{\rm
  eff}(0)=U-\lambda$ ($\lambda=2g^2/\omega_0$), and the sign of this
interaction depends on the relative strength of $U$ and $\lambda$.
In studying the competition between AFM and CO, it is not surprising
to find that the transition between these states occurs when $U\sim \lambda$,
as $U_{\rm eff}(0)>0$ favors AFM and $U_{\rm eff}(0)<0$, the CO state.
What is a surprising result of this study is that this condition still has
some validity in the strong coupling regime, when both $U$ and $\lambda$ are
large, as the transition is still found to occur when $U\sim\lambda$.
We find, however, the nature of the transition does depend on the strength of the couplings,
and also the phonon frequency $\omega_0$. The transition is found to be continuous for
weak 
couplings, and a high phonon frequency $\omega_0$, but becomes discontinuous in the strong
coupling regime, and for smaller values of $\omega_0$.

To gain further insight into this result, we have looked in detail at the
quasiparticle excitations in the normal state. We have calculated both the
quasiparticle weights $z$ and the effective local quasiparticle interaction
$U^r$. We find that the local quasiparticle
interaction $U^r$  changes sign when $U\sim\lambda$, just in the region where the
AFM-CO transition occurs; this is consistent with the
interpretation
of  the transition as due to a Fermi liquid instability. Though the interaction
between the quasiparticle goes to zero at $U\sim \lambda$, the quasiparticles
may still 
be quite significantly renormalized. For example, for $U=\lambda=5$ we find
$z\simeq 0.3$. The fact that the local quasiparticle interaction goes to zero in the
region $U\sim \lambda$, corresponding to $U_{\rm eff}(0)=0$, suggests that the two terms
contributing to  $U_{\rm eff}(\omega)$, are renormalized on the very low
energy scale in a similar way. This is somewhat surprising, as in considering
similar competing interactions in the case of superconductivity, it is
generally
assumed that the dominant renormalization is of the Coulomb term so that it
does not overwhelm the attractive term from phonon exchange. That the two
terms
are renormalized here in a similar way may be a feature of the Holstein-Hubbard 
model, where the phonon term is coupled to the occupation of a local charge.
A model in which the phonons are coupled to a redistribution of the local
charge, as with a coupling to Jahn-Teller modes, might behave differently.
This topic deserves further
investigation.

In calculating the individual contributions to the total energy in the
different ordered states and the normal state, we have been able to show
the subtle interplay of the various terms. These  vary in the weak and
strong coupling regimes, and may change discontinuously at the transition.
 They  also depend on the phonon frequency $\omega_0$.
In the weak coupling regime the energy gain in the broken symmetry state
is via a reduction of the potential energy relative to that of the normal 
state, whereas at strong coupling it is the kinetic energy which is lower in
the  ordered state. This appears to be a general feature.

The final part of this study has been concerned with the spectra, both of the
electrons and phonons. The main effects seen in the phonon spectra are a
softening
and a broadening of the phonon mode in the region of the transition.
In the AFM and CO states well away  from the transition there
is little effect of the coupling to the electrons on the phonon spectrum.
It is more difficult to summarize the results for the electron spectra,
as there  significant differences develop on all energy scales
as the interaction parameters are varied, and as the long range AFM or CO develops.
For relatively small values of $U$ and $\lambda$ the main differences are in the region
near the Fermi level for AFM or CO states compared with the normal state.
This is due to the development of the sublattice structure. For large values
of
$U$ and small values of $\lambda$ there is the triple peak structure of the Hubbard model,
with the narrow renormalized quasiparticle band at the Fermi level flanked by
the broadened 'atomic' peaks. As $\lambda$ is increased to  $\lambda\sim U$,
in the normal state, the narrow quasiparticle band persists, though broadened
somewhat, and the atomic-like peaks broadened into shoulders. In the AFM or CO
states, the quasiparticle band at the Fermi level develops the features
associated with the sublattice structure, as in the weak coupling case.
This is also accompanied by much larger shifts of spectral weight 
on the high energy scales in the sublattice spectral density.

\bigskip
\noindent{\bf Acknowledgment}\par
\noindent
We wish to thank O. Gunnarsson, R. Zeyher for helpful discussions, W. Koller and D. Meyer for their 
earlier contributions to the development of the NRG programs, and F. Trousselet
for critically reading the manuscript.

\begin{appendix}
\section{Mean field theory in the adiabatic limit}
\label{mfapp}
For the mean field theory in the adiabatic limit, the starting point is the
Hamiltonian in the form 
\begin{eqnarray}
H&=&-t \sum_{i,j,{\sigma}}(\elcre i{\sigma}\elann
j{\sigma}+\hc)+U\sum_i\hat n_{i,\uparrow}\hat n_{i,\downarrow}
\label{hubholham2}\\
&&+g_{\rm F}\sum_i \hat
x_i\Big(\sum_{\sigma}\hat n_{i,\sigma}-1\Big)+\sum_i\frac{\hat
  p^2_i}{2M}+\frac{k}2\hat x_i^2, 
\nonumber
\end{eqnarray}
where the parameters of (\ref{hubholham2}) and (\ref{hubholham}) are related
by $\omega_0=~\sqrt{k/M}$, $g_{\rm   F}=\sqrt{2\omega_0}g$, and
$\lambda=g^2_{\rm F}/k$. In this Hamiltonian we can take the limit 
$M\to \infty$, such that the kinetic term for the phonons vanishes and we
replace the operator $\hat x_i$ by a static field $x_i$. From this we obtain
in mean field theory the potential 
\begin{equation}
  V(x_i)=\sum_i\frac{k}2x_i^2+g_{\rm F}\sum_i
  x_i (n_i-1)+E_{\rm kin}+E_{U},
\end{equation}
where $n_i=\sum_{\sigma}\expval{\hat n_{i,\sigma}}{}$.
The condition for a local minimum $\partial V(x_i)/\partial x_i=0$ yields,
\begin{equation}
  x_i=-\frac{g_{\rm F}}k (n_i-1).
\label{xnrel}
\end{equation}
We restrict ourselves to homogeneous solutions, and from the Hamiltonian the
mean field self-energy can be read off,
\begin{equation}
\Sigma_{\alpha,\sigma}(\omega)=Un^{\alpha}_{-\sigma}-\lambda(n^{\alpha}-1),  
\end{equation}
independent of $\omega$ and we have employed (\ref{xnrel}) for the second
term. The index $\alpha=A,B$ corresponds to the sublattice 
and $\sigma$ to the spin. 

In order to determine
$n^{\alpha}_{\sigma}\equiv  \expval{\hat n_{\sigma}^{\alpha}}{}$
we need to consider the equation
\begin{eqnarray*}
  n^{\alpha}_{\sigma}=\frac2N\sum_{\vk}\expval{\elcre{\alpha,\vk}{\sigma}}{\elann{\alpha,\vk}{\sigma}}=
 \frac2N\sum_{\vk}\integral{\omega}{-\infty}{\infty}f_-(\omega)\rho_{\alpha,\vk,\sigma}(\omega),
\end{eqnarray*}
where 
\begin{equation}
  \rho_{\alpha,\vk,\sigma}(\omega)=-\Imag\frac{\zeta_{\bar \alpha,\sigma}(\omega^+)/\pi}
{\zeta_{A,\sigma}(\omega^+)\zeta_{B,\sigma}(\omega^+)-\epsilon_{\vk}^2},
\end{equation}
with $\omega^+=\omega+i\eta$.
We have used the matrix Green's function for the bipartite lattice in the form
(\ref{kgf}). 
This is most easily evaluated with the identity
\begin{equation}
-\Imag\frac{\zeta_{\bar\alpha,\sigma}(\omega^+)/\pi}
{\zeta_{A,\sigma}(\omega^+)\zeta_{B,\sigma}(\omega^+)-\epsilon_{\vk}^2}
=\sum_{m=\pm} u_{m,\sigma}^{\alpha}(\epsilon_{\vk})\delta[\omega-\omega_{m,\sigma}(\epsilon_{\vk})].
\label{deltaid1}
\end{equation}
The excitation $\omega_{m,\sigma}(\epsilon_{\vk})$ are determined from the
poles of the Green's function, 
\begin{equation}
  \zeta_{A,\sigma}(\omega)\zeta_{B,\sigma}(\omega) -\epsilon_{\vk}^2=0,
\end{equation}
which yields generally
\begin{equation}
\omega_{\pm,\sigma}(\epsilon_{\vk})=\frac{\Sigma_{A,\sigma}-\mu_{A,\sigma}+\Sigma_{B,\sigma}-\mu_{B,\sigma}}{2}
\pm E_{{\vk},\sigma},
\end{equation}
where
\begin{equation}
E_{\vk,\sigma}=\sqrt{\epsilon_{\vk}^2+
\frac{[\mu_{A,\sigma}-\Sigma_{A,\sigma}-(\mu_{B,\sigma}-\Sigma_{B,\sigma})]^2}4}.
\end{equation}
The weights $u_{m,\sigma}^{\alpha}(\epsilon_{\vk})$ are generally given
by the inverse of the derivative w.r.t $\omega$ of
\begin{equation}
  f^{\alpha}_{\sigma}(\omega)=\zeta_{\alpha,\sigma}(\omega) -
  \frac{\epsilon_{\vk}^2}{\zeta_{\bar\alpha,\sigma}(\omega)} 
\end{equation}
evaluated at $\omega_{\pm,\sigma}(\epsilon_{\vk})$,
\begin{equation}
  u_{m,\sigma}^{\alpha}(\epsilon_{\vk})=\frac{\zeta_{\bar\alpha,\sigma}(\omega_{m,\sigma}(\epsilon_{\vk}))^2}
{\epsilon_{\vk}^2+\zeta_{\bar\alpha,\sigma}(\omega_{m,\sigma}(\epsilon_{\vk}))^2}.   
\end{equation}
Using these result we find
\begin{equation}
  n^{\alpha}_{\sigma}=\sum_{m}\integral{\epsilon}{}{}
\frac{\rho_0(\epsilon)u_{m,\sigma}^{\alpha}(\epsilon)}{1+\e^{\beta\omega_{m,\sigma}(\epsilon)}},
\end{equation}
through which $n^{\alpha}_{\sigma}$ can be determined self-consistently.

Once $n^{\alpha}_{\sigma}$ is determined we can calculate the ground state
energy to determine which state has the lowest energy. The expression for the total energy reads
\begin{eqnarray*}
E_{\rm mf}&=&
\frac1N\sum_{\vk,\sigma}(\epsilon_{\vk}\expval{\elcre{A,\vk}{\sigma}\elann{B,\vk}{\sigma}}{}+\hc) \\
&&-\frac{\lambda}2 \sum_{\alpha}(n^{\alpha}-1)^2 +\frac
U2\sum_{\alpha} n_{\alpha,\uparrow}n_{\alpha,\downarrow}, 
\end{eqnarray*}
where we have substituted (\ref{xnrel}) for $x$.
This can also be written as
\begin{eqnarray*}
E_{\rm mf}&=&
E_{\rm kin}^{\rm mf}
- \lambda[ (\Phi_{\rm co}^{A})^2+(\Phi_{\rm co}^{B})^2]\\
&&+\frac U2\Big(\frac{n_{A}^2}4-m_A^2+\frac{n_{B}^2}4-m_B^2\Big),
\end{eqnarray*}
where
\begin{eqnarray*}
E_{\rm kin}^{\rm mf}=
\sum_{\sigma}\integral{\epsilon}{}{}
\frac{\rho_0(\epsilon)\epsilon^2}{2E_{\sigma}(\epsilon)}
\Big(\frac{1}{1+\e^{\beta\omega_{+,\sigma}(\epsilon)}}-\frac{1}
{1+\e^{\beta\omega_{-,\sigma}(\epsilon)}}\Big).
\end{eqnarray*}
For half filling, $\Phi_{\rm co}=|\Phi_{\rm co}^{A}|=|\Phi_{\rm co}^{B}|$,
$\Phi_{\rm afm}=|\Phi_{\rm afm}^{A}|=|\Phi_{\rm afm}^{B}|$
this can be written in the simple form
\begin{equation}
  E_{\rm mf}=
 E_{\rm kin}^{\rm mf}
+(U- 2\lambda)\Phi_{\rm co}^2-U\Phi_{\rm afm}^2+\frac U4
\label{Emfsimp}
\end{equation}
From this we can see that if the order parameters are equal and exclusive the
CO state has lower energy for $\lambda>U$ and the AFM state otherwise.

\section{Calculation of the self-energy}
\label{selfencal}
In NRG calculations it is common practice to determine the self-energy from
the Green's function $G_{\alpha,\sigma}(\omega)$ and the higher order Green's
function $F_{\alpha,\sigma}(\omega)$ via \cite{BHP98}
\begin{equation}
\Sigma_{\alpha,\sigma}(\omega)=U\frac{F_{\alpha,\sigma}(\omega)}{G_{\alpha,\sigma}(\omega)}.
\label{sefgf}
\end{equation}
This can be derived in an equations of motion
approach. As $F,G$ are complex functions, $F=F^{\rm R}+iF^{\rm I}$,  we can
write 
\begin{equation}
\Sigma=U\frac{F^{\rm R} G^{\rm R}+F^{\rm I}G^{\rm I}+i(F^{\rm I}G^{\rm
    R}-F^{\rm R}G^{\rm I})}{(G^{\rm R})^2+(G^{\rm I})^2},
\label{sefgf2}
\end{equation}
where we have omitted the indices and the arguments. The procedure
(\ref{sefgf}) for obtaining $\Sigma$ has turned out to 
work well in many cases both for impurity models and lattice models within the
DMFT framework.\cite{BCP08} 
The imaginary part of the retarded self-energy has
the well-known property $\Imag\Sigma_{\alpha,\sigma}(\omega)<0$, which is
respected in Eq. (\ref{sefgf2}) if $F$ and $G$ are the exact Green's
functions. However, in a numerical  self-consistent DMFT calculation of $G,F$
small inaccuracies  - usually near $|\omega|= 0$ -
can lead to $F^{\rm I}(\omega)G^{\rm R}(\omega)-F^{\rm R}(\omega)G^{\rm
  I}(\omega)>0$ and thus slightly positive values for $\Imag\Sigma$ via
Eq. (\ref{sefgf2}). Clearly this is physically incorrect. We have used  
two different ad-hoc procedures to deal  with this complication. The first one (a) is to subtract from  
$\Sigma^{\rm I}_{\alpha,\sigma}(\omega)$ the values by which it exceeds zero
in a certain interval around $\omega=0$. The second one (b)
is to cut-off $\Sigma^{\rm I}_{\alpha,\sigma}(\omega)$ at zero, i.e. to set it
equal to zero for all values where it is positive. We found that
in most cases the procedures give approximately the same result.  However, very close
to the AFM-CO transition the procedure can have an effect on the final result
obtained via the self-consistency equation. One finds that ordered solutions are a bit
less stable for method (a). We have decided to present all results in this paper obtained by
using method (b).

\end{appendix}


\bibliography{artikel,biblio1}

\end{document}